\documentclass[pra,aps,twocolumn,groupedaddress,nofootinbib,superscriptaddress]{revtex4}
\usepackage{graphicx}
\usepackage[nointegrals]{wasysym} \usepackage[export]{adjustbox}
\usepackage{amsmath,amsfonts,amssymb,latexsym}
\usepackage{hhline}
\usepackage{bm}
\usepackage{verbatim}
\usepackage{enumitem}
\usepackage{mathrsfs}
\usepackage{slashed}
\usepackage{empheq}

\newcommand{\mi}{\mathsf{MI}}

\newcommand{\p}{\partial}

\newcommand{\lan}{\langle}
\newcommand{\ran}{\rangle}

\newcommand{\unit}{\mathbf{1}}

\newcommand{\da}{{\dagger}}
\newcommand{\doa}{\downarrow}
\newcommand{\upa}{\uparrow}
\newcommand{\ob}[1]{\mkern 1.5mu\overline{\mkern-1.5mu#1\mkern-1.5mu}\mkern 1.5mu}

\newcommand{\ra}{\rightarrow}

\newcommand{\wt}{\widetilde}

\renewcommand{\(}{\left(}
\renewcommand{\)}{\right)}
\renewcommand{\[}{\left[}
\renewcommand{\]}{\right]}
\newcommand{\mt}{\mapsto}
\newcommand{\xra}{\xrightarrow}

\newcommand{\twp}{{2\pi}}

\newcommand{\D}{\nabla}

\newcommand\bpm            {\begin{pmatrix}}
	\newcommand\epm           {\end{pmatrix}}

\def\app#1#2{	\mathrel{		\setbox0=\hbox{$#1\sim$}		\setbox2=\hbox{			\rlap{\hbox{$#1\propto$}}			\lower1.1\ht0\box0		}		\raise0.25\ht2\box2	}}

\newcommand{\tw}{\textwidth}

\newcommand{\ct}{\Theta}
\newcommand{\vs}{\varsigma}

\newcommand{\inv}{^{-1}}

\newcommand{\ope}\odot

\usepackage{manfnt}

\newcommand{\bi}{\begin{itemize}}
	\newcommand{\ei}{\end{itemize}}

\usepackage{amsthm}

\newtheorem{theorem}{Theorem}

\theoremstyle{definition}
\newtheorem{definition}{Definition}
\theoremstyle{definition}

\newcommand\bd            {\begin{definition}}
	\newcommand\ed            {\end{definition}}
\newcommand\bt            {\begin{theorem}}
	\newcommand\et            {\end{theorem}}
\newcommand\be            {\begin{equation}}
	\newcommand\ee            {\end{equation}}
\newcommand\ba            {\begin{aligned}}
	\newcommand\ea            {\end{aligned}}
\newcommand\bea{\begin{equation}\begin{aligned}}
		\newcommand\eea{\end{aligned}\end{equation}}

\usepackage{subcaption}
\usepackage[usenames,dvipsnames]{xcolor}

\newcommand{\ethan}[1]{ { \color{red} \footnotesize \textsf{EL: \textsl{#1}} }}

\usepackage{hyperref} 
\hypersetup{final}
\hypersetup{colorlinks, citecolor=red, linkcolor=red, urlcolor=red}

\renewcommand{\ss}{\subsection}

\renewcommand{\a}{\alpha}
\renewcommand{\b}{\beta}
\renewcommand{\d}{\delta}
\newcommand{\De}{\Delta}
\newcommand{\g}{\gamma}

\newcommand{\s}{\sigma}

\newcommand{\ep}{\varepsilon} \renewcommand{\l}{\lambda}
\renewcommand{\L}{\Lambda}
\renewcommand{\t}{\theta}

\renewcommand{\o}{\omega}

\renewcommand{\r}{\rho}

\renewcommand{\c}{\chi}

\newcommand{\nn}{\mathbb{N}}

\newcommand{\rr}{\mathbb{R}}
 \newcommand{\qq}{\qquad}

\newcommand{\zz}{\mathbb{Z}}

\newcommand{\mco}{\mathcal{O}}

\newcommand{\mcj}{\mathcal{J}}

\newcommand{\mcl}{\mathcal{L}}

\newcommand{\mct}{\mathcal{T}}
\newcommand{\mch}{\mathcal{H}}
\newcommand{\mca}{\mathcal{A}}
\newcommand{\mcp}{\mathcal{P}}

\newcommand{\mcz}{\mathcal{Z}}

\newcommand{\sfc}{\mathsf{c}}

\newcommand{\sfm}{\mathsf{m}}

\usepackage[mathscr]{eucal}

\usepackage{braket}

\renewcommand{\k}{\ket}

\usepackage{dcolumn}
\usepackage{pifont}
\usepackage{ulem}

\begin{document}
	
	\title{Dipole condensates in tilted Bose-Hubbard chains}

	\author{Ethan Lake}
	\affiliation{Department of Physics, Massachusetts Institute of Technology, Cambridge, MA, 02139}
	\author{Hyun-Yong Lee}
	\affiliation{Division of Display and Semiconductor Physics, Korea University, Sejong 30019, Korea}
	\affiliation{Department of Applied Physics, Graduate School, Korea University, Sejong 30019, Korea}
	\affiliation{Interdisciplinary Program in E·ICT-Culture-Sports Convergence, Korea University, Sejong 30019, Korea}
	\author{Jung Hoon Han}
	\affiliation{Department of Physics, Sungkyunkwan University, Suwon 16419, Korea}
	\author{T. Senthil} 
	\affiliation{Department of Physics, Massachusetts Institute of Technology, Cambridge, MA, 02139}

	\begin{abstract}
		We study the quantum phase diagram of a Bose-Hubbard chain whose dynamics conserves both boson number and boson dipole moment, a situation which can arise in strongly tilted optical lattices. The conservation of dipole moment has a dramatic effect on the phase diagram, which we analyze by combining a field theory analysis with DMRG simulations. In the thermodynamic limit, the phase diagram is dominated by various types of incompressible dipolar condensates. In finite-sized systems however, it may be possible to stabilize a `Bose-Einstein insulator': an exotic compressible phase which is insulating, despite the absence of a charge gap. 
		We suggest several ways by which these exotic phases can be identified in near-term cold atom experiments. 
	\end{abstract}
	
	\maketitle

	\section{Introduction and summary} 
	
	Many of the most fascinating phenomena in quantum condensed matter physics arise from the competition between kinetic energy and interactions, and it is therefore interesting to examine situations in which the roles played by either kinetic energy or interactions can be altered. One way of doing this is by finding a way to quench the system's kinetic energy. This can be done with strong magnetic fields---which allows one to explore the rich landscape of quantum Hall phenomenology---or by engineering the system to have anomalously flat energy bands, as has been brought to the forefront of condensed matter physics with the emergence of Moire materials \cite{andrei2021marvels}. 
	
	A comparatively less well understood way to quench kinetic energy occurs when exotic conservation laws inhibit particle motion. One large class of models in which this mechanism is operative are systems whose dynamics conserves the {\it dipole moment} ($i.e.$ center of mass) of the system's constituent particles, in addition to total particle number \cite{pretko2017subdimensional,pretko2018fracton}. This conservation law can be easily engineered as an emergent symmetry in strongly-tilted optical lattices, where energy conservation facilitates dipole-conserving dynamics over arbitrarily long pre-thermal timescales \cite{guardado2020subdiffusion,scherg2021observing,kohlert2021experimental} (other physical realizations are discussed below). 
	
\begin{figure} 
		\includegraphics[width=.5\tw]{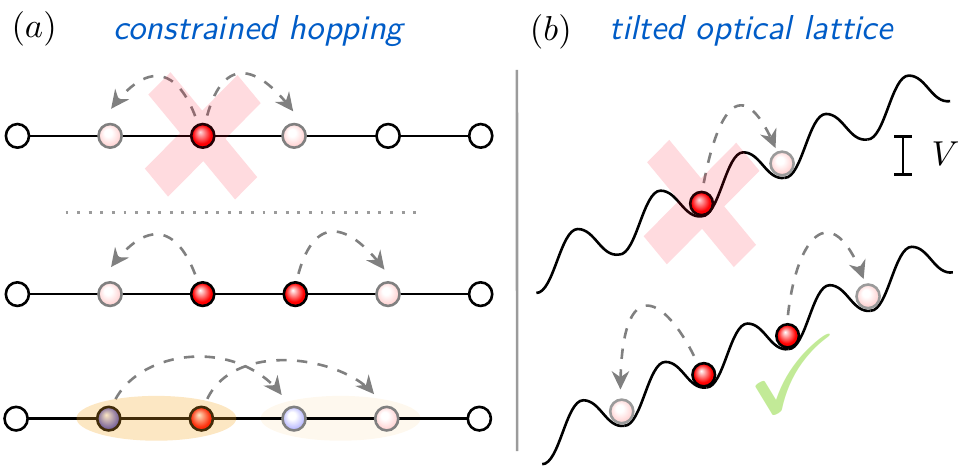}
		\caption{\label{fig:inspir_schem}  (a) The restricted kinematics of dipole conservation. An isolated boson cannot move (top), while two nearby bosons can move only by coordinated hopping in opposite directions (middle). A boson and a hole (blue circle) can move freely in both directions (bottom). (b) Approximate dipole conservation can be engineered in tilted optical lattices with large tilt strength $V$. Energy conservation then forbids single bosons from hopping (top), while dipole-conserving hopping processes are allowed (bottom).
		} 
	\end{figure}

	Dipole conservation prevents individual particles from moving independently on their own [Fig. \ref{fig:inspir_schem} (a)]. Instead, motion is possible in only one of two ways: first, two nearby particles can `push' off of each other, and move in opposite directions. This type of motion allows particles to hop over short distances, since this process freezes out as the particles get far apart. Second, a particle and a hole (where `hole' is defined with respect to a background density of particles) can team up to form a dipolar bound state, which --- by virtue of the fact that it is charge neutral --- may actually move freely, without constraints. Dipole conservation thus forces the system's `kinetic energy' to be {\it intrinsically nonlinear}, as the way in which any given particle moves is always conditioned on the charge distribution in its immediate vicinity. This leads to a blurring of the lines between kinetic energy and interactions, producing a wide range of interesting physical phenomena. 
	
	\begin{figure*}
		\includegraphics[width=\tw]{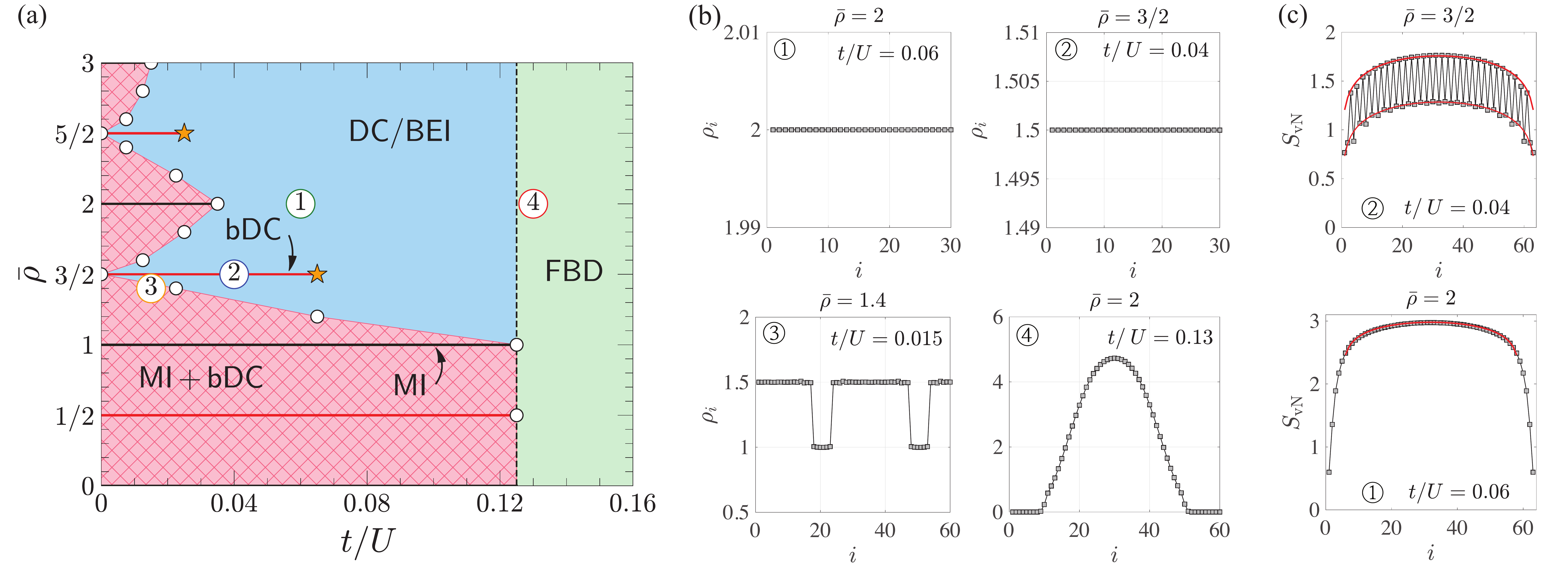}
		\caption{\label{fig:dmrg} 
			(a) thermodynamic phase diagram of the 1d DBHM as a function of boson filling $\ob \r$ and hopping strength $t/U$ (with $t_4=t_3\equiv t$). The blue region denotes a dipole condensate (DC) which for non-integer $\ob\r$ has CDW order. {In finite-sized systems the DC may give way to a Bose-Einstein insulator, although where exactly this is most likely to occur is non-universal.} Red lines denote a different type of DC marked as `bDC' (here `b' stands for {\it bond}-centered CDW; see text for details), whose existence relies on having a nonzero $t_4$.
			Black lines at integer $\ob\r$ denote Mott insulators, and in the shaded pink region phase separation between the MI and bDC phases occurs. The green region at largest $t/U$ denotes the fractured Bose droplet (FBD) phase. The small white circles signify the location of the phase boundary as obtained in DMRG.
			(b) Plots of the real-space average density $\lan\rho_i\ran$ for the four points marked on the phase diagram. (c) Entanglement entropy in the DC phase (top) and the bDC (bottom). The red lines are $c=1$ fits to the Calabrese-Cardy formula for the entanglement entropy $S_i = (c/6) \log [(2L/\pi) \sin (\pi i /L )]$ for a finite chain of length $L$~\cite{calabrese2009entanglement}. 
		} 
	\end{figure*}

	The effects of dipole conservation on quantum dynamics have been explored quite intensely in recent years, with the attendant kinetic constraints often leading to Hilbert space fragmentation, slow thermalization, and anomalous diffusion (e.g. \cite{khemani2020localization,sala2020ergodicity,pai2019localization,rakovszky2020statistical,moudgalya2019thermalization,van2019bloch,schulz2019stark,gromov2020fracton,feldmeier2020anomalous,iaconis2021multipole,glorioso2021breakdown,grosvenor2021hydrodynamics,radzihovsky2020quantum,moudgalya2022hilbert}). On the other hand, there has been comparatively little focus on understanding the {\it quantum ground states} of dipole conserving models \cite{lake2022dipolar,prem2018emergent,yuan2020fractonic,chen2021fractonic}. Addressing this problem requires developing intuition for how interactions compete with an intrinsically nonlinear form of kinetic energy, and understanding the types of states favored when the dipolar kinetic energy dominates the physics. 
	
	One concrete step towards addressing these questions was given in Ref. \cite{lake2022dipolar}, which put forward the {\it dipolar Bose-Hubbard model} (DBHM) as a representative model that succinctly captures the effects of dipole conservation. The DBHM is simply a dipole-conserving version of the well-loved Bose Hubbard model, and displays a variety of interesting phases with rather perplexing properties, all driven by the physics of the dipolar bound states mentioned above. Of particular interest is the {\it Bose-Einstein insulator} (BEI) phase identified in Ref. \cite{lake2022dipolar}, which is realized in the regime where the nonlinear kinetic energy dominates. The BEI is compressible, and contains a Bose-Einstein condensate, but remarkably is nevertheless {\it insulating}, and has vanishing superfluid weight. 
	
	The aim of the present work is to perform a detailed investigation of the DBHM in one dimension, with the aim of fully understanding the phase diagram and making concrete predictions for near-term cold atom experiments. We are able to understand the entire phase diagram within a concise field theory framework, whose predictions we confirm with extensive DMRG simulations. {We will see that the physics of the DBHM in 1d is slightly different from the 2d and 3d versions, in that the BEI phase is absent in the thermodynamic limit, being rendered unstable by a particular type of relevant perturbation. However, if the bare strength of these destabilizing perturbations is very small it may be possible to stabilize a BEI regime up to (potentially very large) length scales. In the following we will see various pieces of numerical evidence that this indeed can occur at fractional fillings. }
	
	The concrete model we will study in this paper is a modified version of the standard Bose-Hubbard model, with the hopping terms modified to take into account dipole conservation:
	\bea \label{dbhm_ham} H_{DBHM} & = - \sum_i \(t_3 b_{i-1}^\da b_i^2 b_{i+1}^\da + t_4 b^\da_{i-1} b_i b_{i+1} b_{i+2}^\da + h.c.\) \\ &  + \frac U2\sum_i  n_i ^2,\eea
	where $n_i = b^\da_i b_i$ is the boson number operator on site $i$, and $t_{3,4}\geq0$ determine the strength of the dipole-conserving hopping processes (see Sec. \ref{sec:realizations} for an explanation of how $H_{DBHM}$ arises in the tilted optical lattice context). 
	In this paper we will always work in the canonical ensemble, with the boson density 
	fixed at 
	\be \ob \r = \frac{n}{m},\qq {\rm gcd}(n,m) = 1,\ee 
	where we are working in units where the lattice spacing is equal to unity. Our goal is to study the behavior of the ground states of $H_{DBHM}$ as a function of $t_{3,4}/U$ and $\ob\r$.\footnote{See also Ref. \cite{rakovszky2020statistical} for a discussion of the ground state physics of a dipolar spin-1 model, which in some aspects behaves similarly to our model at half-odd-integer filling and small $t_{3,4}/U$.}
	
	Let us now give a brief overview of our results. The phase diagram we obtain is shown in Fig. \ref{fig:dmrg} (a), and contains a lot of information. At the present juncture we will only mention its most salient features, leaving a more detailed treatment to the following sections. 
	
	Our first result is that---at least {\it in the thermodynamic limit}---there is a nonzero charge gap to exciting single bosons {\it throughout} the phase diagram. Regardless of $t/U$ and $\ob\r$, the boson correlation functions always decay as $\lan b_i^\da b_j\ran \sim e^{-|i-j|/\xi}$, with the ground state being incompressible across the entire phase diagram. This is rather remarkable, as the system thus remains incompressible over a {\it continuous} range of filling fractions, and moreover does so without disorder, and with only short-ranged interactions. It manages to do this by having vortices in the boson phase condense at all points of the phase diagram, a situation which is made possible by the way in which dipole conservation modifies vortex energetics. This is of course in marked contrast to the regular Bose-Hubbard model, which has a compressible superfluid phase which is broken by incompressible Mott insulators only at integer filling fractions and weak hopping strengths \cite{fisher1989boson}. 
	
	{While the statements in the above paragraph are correct in the thermodynamic limit, the ubiquitous vortex condensation just discussed may be suppressed in finite systems. This will happen if the operators which create vortices can take a long time to be generated under RG, becoming appreciable only at extremely large length scales. In our model this is particularly relevant at fractional fillings, where a BEI seems to be realized in our DMRG numerics. In Sec. \ref{sec:bei} we study this phenomenon from a different point of view within the context of a dipolar rotor model. }
	
	
	The broad-strokes picture of our phase diagram is dictated by the physics of neutral `excitonic' dipolar bound states of particles and holes annihilated by the dipole operators
	\be d_i \equiv b^\da_i b_{i+1}.\ee 
	As discussed above, the motion of these dipolar particle-hole bound states is not constrained by dipole conservation. This can be seen mathematically by noting that the hopping terms in $H_{DBHM}$ can be written using the $d_i$ operators as 
	\be H_{hop} = -\sum_i (t_3d_i^\da d_{i+1} + t_4d_i^\da d_{i+2} + h.c.),\ee 
	and thus constitute {\it conventional} hopping terms for the dipolar bound states.  Since the dipolar bound states are allowed to move freely, it is natural to expect that the most efficient way for the system to lower its energy is for them to Bose condense. Indeed, this expectation is born out in both our field theory analysis and in our numerical simulations. The result of this condensation is an exotic gapless phase we refer to as a {\it dipole condensate} (DC). Unlike the superfluid that occurs in the standard Bose-Hubbard model, the dipole condensate realized here is formed by charge {\it neutral} objects, and in fact actually has {\it vanishing} DC conductivity. 
	
	A second interesting feature of the DBHM is an instability to an exotic glassy phase we dub the {\it fractured Bose droplet} (FBD) phase. This instability occurs in the green region drawn in the phase diagram of Fig. \ref{fig:dmrg} (a), and is in fact common to all dipole-conserving boson models of the form \eqref{dbhm_ham}. It arises simply due to the $\sqrt{n}$ factors appearing in $b\k n = \sqrt n \k{n-1}, b^\da \k{n-1} = \sqrt n \k n$. These factors mean that when acting on a state of average density $\ob\r$, the dipolar hopping terms in the Hamiltonian scale as $-2(t_3+t_4)\ob\r^2$ at large $\ob\r$, precisely in the same way as the Hubbard repulsion term, which goes as $+\frac12U\ob\r^2$. Thus when 
	\be\label{fbd_cond} t_3+t_4 \geq  \frac{U}{4},\ee 
	it is always energetically favorable to locally make $\ob\r$ as large as possible (in our phase diagram, DMRG finds an instability {\it exactly} when this condition is satisfied). Once this occurs, the lowest-energy state of the system will be one in which all of the bosons agglomerate into one macroscopic droplet [Fig. \ref{fig:dmrg} (b), panel 4].

	The physics of the FBD phase is actually much more interesting than the above discussion might suggest: rather than simply forming a giant droplet containing an extensively large number of bosons, dipole conservation means that the system instead fractures into an interesting type of metastable glassy state (the physics of which may underlie the observation of spontaneous formation in the dipole-conserving 2d system of Ref. \cite{zahn2022formation}). Understanding the physics of this phase requires a set of theoretical tools better adapted to addressing dynamical questions, and will be addressed in a separate upcoming work \cite{fbd} (see also the fractonic microemulsions of Ref. \cite{prem2018emergent}). 
	
	The remainder of this paper is structured as follows. 
	In the next section (Sec. \ref{sec:realizations}), we briefly discuss various possible routes to realizing the DBHM in experiment, focusing in particular on the setup of tilted optical lattices. In Sec.	
	\ref{sec:field_theory} we develop a general field theory approach that we use to understand the phase diagram in broad strokes. This approach in particular allows us to understand the role that vortices in the boson phase play in determining the nature of the phase diagram. In Sec. \ref{sec:expt} we discuss a few characteristic features possessed by the dipole condensate, as well as how to detect its existence in experiment. In Sec. \ref{sec:bei}, we give a more detailed discussion of how the exotic physics of the Bose-Einstein insulator may appear in small systems due to finite-size effects. 
	The following sections \ref{sec:int_filling}, \ref{sec:half_filling}, and \ref{sec:gen_filling} discuss in detail the physics at integer, half-odd-integer, and generic filling fractions, respectively. We conclude with a summary and outlook in Sec. \ref{sec:disc}. 
	
	\section{Experimental realizations} \label{sec:realizations} 
	
	Before discussing the physics of the DBHM Hamiltonian \eqref{dbhm_ham} in detail, we first briefly discuss pathways for realizing dipole conserving dynamics in experiment. 
	
	The simplest and best-explored way of engineering a dipole conserving model is to realize $H_{DBHM}$ as an effective model that describes the prethermal dynamics of bosons in a strongly tilted optical lattice \cite{khemani2020localization,scherg2021observing,guardado2020subdiffusion,sala2020ergodicity}. In this context, the microscopic Hamiltonian one starts with is
	\bea H_{tilted} & = -t_{sp} \sum_i (b_i^\da b_{i+1} +b_{i+1}^\da b_i)+ \sum_i V i n_i + H_U,\eea 
	where $H_U$ denotes the Hubbard repulsion term, and $V$ is the strength of the tilt potential (which in practice is created with a magnetic field gradient). In the strong tilt limit where $t_{sp}/V,U/V\ll1$,\footnote{See e.g. \cite{sachdev2002mott,pielawa2011correlated,yang2020observation,su2022observation} for discussions of tilted Bose-Hubbard models in regimes with weaker $V$.} energy conservation prevents bosons from hopping freely, but does not forbid coordinated hopping processes that leave the total boson dipole moment invariant [Fig. \ref{fig:inspir_schem} (b)]. Perturbation theory to third order \cite{scherg2021observing,moudgalya2019thermalization} then produces the dipolar model \eqref{dbhm_ham} with $t_3= t_{sp}^2 U / V^2, t_4 = 0$, and an additional nearest-neighbor interaction $ (2t^2/V^2)\sum_i n_in_{i+1}$---see App. \ref{app:hopping} for the details. A nonzero $t_4$ will eventually be generated at sixth order in perturbation theory (or at third order, if one adds an additional nearest-neighbor Hubbard repulsion), but in the optical lattice context we generically expect $t_4/t_3 \ll 1$. We note however that the DMRG simulations that we discuss below are performed with a nonzero $t_4$ (in fact for simplicity, we simply set $t_4 = t_3$). This is done both because one can imagine other physical contexts in which an appreciable $t_4$ coupling is present, and because the $t_4$ term moderately helps DMRG convergence. In any case, the qualitative features of the $t_4 = 0$ and $t_3= t_4$ models are largely the same, with the only differences arising near certain phase transitions, and at certain filling fractions (as will be discussed in Sec. \ref{sec:half_filling}). 
	
	An interesting aspect of the effective dipolar Hamiltonian that arises in this setup is that $t_{3} / U$ always scales as $(t_{sp}/V)^2 \ll 1$ \cite{taylor2020experimental}. One may worry that this could lead to problems when trying to explore the full phase diagram, since we will be unable to access regimes in which $t_3 / U \gtrsim 1$. This however does not appear to be a deal-breaker, since all of the action in the DBHM will turn out to occur at $t_3 / U \lesssim 0.1$ (and at large fillings, the dipole condensate in particular turns out to be realizable at arbitrarily small $t_3/U$). 
	
	Another possible realization of the 1d DBHM is in bosonic quantum processors based on superconducting resonators \cite{underwood2012low,wang2020efficient,ma2019dissipatively}, where the dipolar hopping terms can be engineered directly, and there are no fundamental constraints on $t_{3,4}/U$. 
	In this setup there is no way to forbid single particle hopping terms on symmetry grounds alone, and generically the Hamiltonian will contain a term of the form
	\be  \label{hsp} H_{sp}= -t_0 \sum_i ( b^\dag_{i+1} b_i + b^\dag_i b_{i+1}).\ee  
	The presence of such dipole-violating terms is actually not a deal-breaker, as long as $t_0$ is sufficiently small compared to $t_{3,4}$. Indeed, the fact that single bosons are gapped throughout the entire phase diagram means that a sufficiently small $H_{sp}$ will always be unimportant, a conclusion that we verify in DMRG.

	\section{Master field theory} \label{sec:field_theory} 
	
	In the remainder of the main text, we will fix 
	\be t_3 = t_4 \equiv t \ee
	for concreteness, which matches the choice made in the numerics discussed below. Those places where setting $t_4 = 0$ qualitatively changes the physics will be mentioned explicitly.

	In this section we discuss a continuum field theory approach that we will use in later sections as a guide to understand the phase diagram. Our field theory involves two fields $\t$ and $\phi$, which capture the long-wavelength fluctuations of the density and phase, respectively. In terms of these fields, the boson operator is 
	\be\label{brep} b = \sqrt{\r} e^{i\phi},\qq \r = \ob\r + \frac1\twp \p_x^2 \t,\ee
	Note that density fluctuations are expressed as the {\it double} derivative of $\t$ (in the standard treatment \cite{haldane1981effective} there is only a single derivative). This gives the commutation relations 
	\be \label{comms} [\phi(x),\p_y^2\t(y)] = \twp i \d(x-y).\ee 
	The reason for writing the fluctuations in the density in this way will become clear shortly. 
	
	Before discussing how to construct our field theory, let us discuss how $\phi,\t$ transform under the relevant symmetries at play. 
	Dipole symmetry leaves $\t$ alone, but acts as a coordinate-dependent shift of $\phi$, mapping $U(1)_D : \phi(x) \mt \phi(x) + \l x$ for constant $\l$. Thus $e^{i\p_x\phi}$ is an order parameter for the dipole symmetry, since 
	\be U(1)_D : e^{i\p_x\phi} \mt e^{i \l} e^{i\p_x\phi}.\ee 
	
	The operators $e^{i\p_x \t}$, $e^{i\t}$ create vortices\footnote{Since we are in 1d it is more correct to use the word `instanton', but we will stick to `vortex' throughout.} in the phase $\phi$ and its gradient $\p_x\phi$ respectively, which can be shown using the commutation relation \eqref{comms}. Vortices in $\p_x\phi$ are not necessarily objects that we are used to dealing with, but indeed they are well-defined on the lattice \cite{ma2018higher}, and are the natural textures to consider in a continuum limit where $\p_x\phi$ becomes smooth but $\phi$ does not (a limit that dipole symmetry forces us to consider, as this turns out to be relevant for describing the dipole condensate). 
	
	In a background of charge density $\ob \r$ vortices carry momentum $\twp\ob \r$, and so a translation through a distance $\d$ acts as $T_\d : e^{i\p_x \t} \mt e^{i\twp \d n/m} e^{i\p_x \t}$ (recall that $\ob \r = n/m$). 
	To understand this, consider moving a vortex created by $e^{i\p_x\t(x)}$ through a distance $\d$ to the right. Doing so passes the vortex over an amount of charge equal to $\ob\r\d$, which in our continuum notation is created by an operator proportional to $e^{i\d\ob\r \phi(x)}$. Since $e^{i\p_x \t(x)} e^{i\d\ob\r \phi(x)} = e^{i\twp\d\ob\r} e^{i\d\ob\r\phi(x)} e^{i\p_x\t(x)}$, a phase of $e^{i\twp \d\ob \r}$ is accumulated during this process. Consistent with this, a more careful analysis in App. \ref{app:duality} shows that
	\be \label{theta_translation} T_\d : \t(x) \mt \t(x+\d) +\twp \ob \r x\d.\ee 
	
	For our discussion of the phases that occur at fractional fillings, we will also need to discuss how $\t$ transforms under both site- and bond-centered reflections $R_s$ and $R_b=T_{1/2} R_s T_{1/2}$. Using \eqref{theta_translation} and the fact that $R_s : \r(x) \mt \r(-x)$, we see that 
	\bea \label{refl} & R_s: \t(x) \mt \t(-x), \\ & R_b : \t(x) \mt \t(-x)-\frac\pi2\ob\r .\eea 
	
	We now need to understand how to write down a field theory in terms of $\phi$ and $\t$ which faithfully captures the physics of $H_{DBHM}$. The most naive approach is to rewrite $H_{DBHM}$ as 
	\bea 
	\label{expanded_h_mt}
	& H_{DBHM}  = t \sum_i ( | b_{i+1}b_{i-1} - b_i^2 |^2 + | b_{i+2}b_{i-1} - b_{i} b_{i+1} |^2)  \\ 
	& \quad + \sum_i \((U/2-t)n_i^2 - t (n_in_{i+1}+n_in_{i+2} + n_in_{i+3})\),
	\eea 
	and to then perform a gradient expansion. Using the representation \eqref{brep} and keeping the lowest order derivatives of $\t$ and $\phi$, this produces a continuum theory with Hamiltonian density 
	\be \label{mch} \mch = \frac{K_D}2(\p_x^2\phi)^2   + \frac u2 (\p_x^2\t)^2,\ee 
	where we have defined the dipolar phase stiffness $K_D$ and charge stiffness $u$ as 
	\be \label{kd_and_v} K_D \equiv 4\ob \r^2t,\quad u \equiv \frac{U-8t}{(\twp)^2}.\ee
	Taking the above Hamiltonian density $\mch$ as a starting point and integrating out $\t$ produces the Lagrangian of the quantum Lifshitz model studied in Refs. \cite{lake2022dipolar,gorantla2022global}, which describes the BEI phase:
	\be \label{lbei} \mcl_{BEI} = \frac{ K_\tau}2 (\p_\tau \phi )^2 + \frac{K_D}2 (\p_x^2 \phi)^2 ,\ee 
	where $K_\tau \equiv 1/(8\pi^2 u)$. 
	
	The steps leading to \eqref{lbei} miss an essential part of the physics, since they neglect vortices in the phase $\phi$ (as well as vortices in the dipole phase $\p_x\phi$). In the regular Bose-Hubbard model, vortices can be accounted for using the hydrodyanmic prescription introduced by Haldane in Ref. \cite{haldane1981effective}. Using our representation of the density fluctuations as $\p_x^2\t/\twp$, a naive application of this approach would lead to a Lagrangian containing cosines of the form $\cos(l\p_x\t)$, $l\in \nn$. This however turns out to {\it not} fully account for the effects of vortices in the DBHM, which require that the terms $\cos(l\t)$ be added as well.
	The exact perscription for including vortices is worked out carefully in App. \ref{app:duality} using lattice duality, wherein we derive the effective Lagrangian 
	\bea \label{mcl} \mcl_{DBHM} & = \frac{i}{\twp} \p_\tau\phi (\twp \ob \r + \p_x^2\t) +  \frac{K_D}2 (\p_x^2\phi)^2+ \frac{u}{2}(\p_x^2\t)^2 \\ & \qq - y_{D,{4m}} \cos(4m\t)- y_m \cos(m \p_x \t),\eea 
	where the coupling constants $y_l, y_{D,l}$ are given by the $l$-fold vortex and dipole vortex fugacities 
	\be \label{ys} y_l \sim e^{-l^2 c\sqrt{K_D/u}},\qq y_{D,l} \sim e^{-l^2 c_D \sqrt{K_D/u}},\ee
	where $c,c_D$ are non-universal $O(1)$ constants (App. \ref{app:duality} contains the derivation). 
	The appearance of $m$ in the term $y_m\cos(m\p_x\t)$ is due to \eqref{theta_translation}, which ensures that the leading translation-invariant interactions are those which create $m$-fold vortices (recall $\ob\r = n/m$). The factor of $4$ in $y_{D,4m}\cos(4m\t)$ is due to the bond-centered reflection symmetry $R_b$ which shifts $\t$ according to \eqref{refl} (with $\cos(m\t)$ being the most relevant cosine of $\t$ in the absence of $R_b$ symmetry).  
	
	From the above expression \eqref{kd_and_v} for $u$, we see that an instability occurs when 
	\be \label{fbd_inst} t > t_{FBD} \equiv \frac U8,\ee 
	which is precisely the condition given earlier in \eqref{fbd_cond}. 
	When $t>t_{FBD}$, $u$ becomes {\it negative}, and the system is unstable against large density fluctuations---this leads to the glassy phase discussed in the introduction. In the rest of this paper, we will restrict our attention to values of $t$ for which $u>0$, where the above field theory description is valid.
	
	To understand the physics contained in the Lagrangian $\mcl_{DBHM}$, the first order of business is to evaluate the importance of the cosines appearing therein. It is easy to check that at the free fixed point given by the quadratic terms in $\mcl_{DBHM}$ (the first line of \eqref{mcl}), $\cos(l\t)$ has ultra short-ranged correlations in both space and time, for any $l\in\zz$. This is however {\it not} true for $\cos(l\p_x\t)$, whose correlation functions are {\it constant} at long distances, regardless of $l$. 
	This means that $\cos(m\p_x\t)$ is {\it always} relevant, implying that $\p_x\t$ will always pick up an expectation value in the thermodynamic limit, and that vortices will condense at all rational fillings. This is physically quite reasonable due to dipole symmetry forbidding a $(\p_x\phi)^2$ term in \eqref{mcl}, implying that vortices in $\phi$ do not come with the usual logarithmically-divergent gradient energy.
	
	{
		Strictly speaking, this ubiquitous vortex condensation thus prevents the existence of a phase in which the low energy physics is dictated solely by the phase field $\phi$, and consequently preempts the BEI phase (which in the thermodynamic limit can only be realized in $d>1$ spatial dimensions).\footnote{This fact actually has an avatar in 2d classical elasticity theory, where it shows up as the instability of smectics towards nematics. Indeed, integrating out $\phi$ in \eqref{mcl} yields an exact analogue of the Lagrangian describing a 2d smectic \cite{zhai2021fractonic}, with $\cos(\p_x\t)$ the operator sourcing dislocations.} That said, if the vortex fugacity $y_m$ is extremely small (as is likely at filling fractions with large $m$), then the destabilizing cosine $y_m\cos(m\p_x\t)$ will be important only at large distances, leading to a BEI regime emerging on intermediate length scales. As we discuss in Sec. \ref{sec:half_filling}, there is evidence for this occurring in our DMRG numerics at fractional filling, while in Sec. \ref{sec:bei} this is shown to occur in a rotor model that mimics the physics of $H_{DBHM}$ at large densities. }
	
	We now consider what happens when the strength of the $\cos(m\p_x\t)$ term flows to become large enough to impact the low-energy physics. Expanding $\cos(m\p_x\t)$ to quadratic order and integrating out $\phi$, we arrive at the Lagrangian
	\bea \label{mcl_dc} \mcl_{DC} & = \frac1{8\pi^2 K_D} (\p_\tau\t)^2 + \frac{m^2 y_m}2 (\p_x\t-\lan \p_x \t\ran)^2 \\ & \qq - y_{D,4m} \cos(4m\t). \eea 
	The scaling dimension of the remaining cosine is 
	\be \De_{\cos(4m\t)} =  8\sqrt{\frac{m K_D}{y_m}}.\ee  
	When $\De_{\cos(4m\t)} > 2$ this cosine can be dropped, leading to a free quadratic theory for $\t$. This theory describes the {\it dipole condensate} (DC) mentioned in the introduction. Indeed, in this phase dipolar bound states condense and exhibit quasi-long range order (QLRO), with $e^{i\p_x\phi}$ correlators decaying algebraically. On the other hand, individual bosons remain gapped, and the charge compressibility vanishes (more details will be given in Sec. \ref{sec:expt}). 
	When $\De_{\cos(4m\t)} < 2$ on the other hand, $\cos(4m\t)$ is relevant, and $\t$ acquires an expectation value. This consequently proliferates vortices in $\p_x \phi$, destroying the DC and leading to a gapped phase. 
	

	\section{Signatures of the dipole condensate} \label{sec:expt} 
	
	Before embarking on a more detailed tour of the phase diagram, we first briefly discuss the physical properties of the DC, and how it might be detected in near-term experiments on tilted optical lattices. 
	
	We start with the claim made at the beginning of our tour of the phase diagram, namely that single bosons are gapped in the DC, and that the DC---despite being gapless---is in fact an incompressible insulator. We are now in a position to back this up, by calculating correlation functions of $b \sim e^{i\phi}$. 
	Following the procedure outlined in App. \ref{app:duality}, one can show that the IR correlation functions of $e^{i\phi}$, which we write as $C_{e^{i\phi}}(\tau,x) \equiv \langle e^{i\phi(\tau,x)} e^{-i\phi(0,0)}\ran$, are 
	\be \label{phiphi_main}   \ln C_{e^{i\phi}}(\tau,x) =-\int_{q,\o}   \frac{q^2 y_m(1-\cos(qx-\o\tau))}{(\o^2+q^2)^2(\o^2 + 4\pi^2 q^2y_m K_D)}.\ee 
	Just from dimension counting, we see that the integral is IR divergent for all nonzero $\tau,x$, and as such $e^{i\phi}$ correlators are {\it ultralocal} in spacetime. For example, when $x=0$ we obtain 
	\be \label{local_boson_corr} \ln C_{e^{i\phi}}(\tau,0)  = - \frac1{4(1+\vs)^2} \int_\o \frac{1-\cos(\o\tau)}{| \o|^3 }, \ee
	where we have defined $\vs \equiv\twp \sqrt{y_m K_D}$. 
	This integral diverges logarithmically even as $\tau \ra 0$, so that the boson correlation functions are indeed ultralocal, and single bosons are gapped.
	
	Next we consider correlation functions of $d_i\sim e^{i\p_x \phi}$, the dipole order parameter. At equal times, we find 
	\bea \label{cdphi}\ln C_{e^{i\p_x\phi}}(0,x) &= \frac{2+\vs}{4\vs(1+\vs)^2} \int_q \frac{1-\cos(qx)}{|q|}\\ & \ra  \frac{2+\vs}{8\pi\vs(1+\vs)^2} \log(x),  \eea 
	so that $e^{i\p_x\phi}$ has power law correlations with a non-universal exponent depending on $\vs$, with the dipole order parameter thus exhibiting QLRO:
	\be \lan d_i^\da d_j \ran \sim |i-j|^{-\a},\ee 
	with $\a$ a non-universal Luttinger parameter varying continuously within the DC phase. Since the effective IR theory for the DC has dynamical exponent $z=1$, correlations in time behave similarly, as do correlation functions of $e^{i\partial_\tau\phi}$. 
	
	The density-density response is obtained simply from correlation functions of $\p_x^2\t$, yielding 
	\be  \chi_{\r\r}(\o,q) = \frac{q^4}{\o^2/(4\pi^2K_D) + q^2/y_m + m_D^2}, \label{eq:chi-rr} \ee
	where we have allowed for a nonzero effective dipole mass $m_D$, which vanishes when dipoles condense and is nonzero otherwise.
	At small $q$, the charge compressibility thus vanishes as 
	\be \kappa \equiv \chi_{\r\r}(\o,q) |_{\o=0,q\ra 0} = \begin{cases} y_m q^2 & {\rm DC} \\ q^4/m_D^2 & {\rm else}\end{cases}. \label{eq:kappa} \ee 
	The equal-time density-density correlation function $\c(q)\equiv \c_{\r\r}(t=0,q)$ obtained from \eqref{eq:chi-rr} goes as 
	\be \label{chi_equal_time} \c(q) \propto \begin{dcases} |q|^3 & {\rm DC} \\ q^4 & {\rm else} \end{dcases} .\ee 
	Finally, dipole symmetry ensures that the DC conductivity vanishes \cite{lake2022dipolar}, so that the system always is insulating. 
	
	Given the above, what is the best pathway for detecting the DC phase in experiment? This question is slightly subtle, since as we have shown, the DC is an incompressible insulator. One approach would be to directly measure the density-density response function. From the above expression for $\kappa$, this however requires resolving the difference between $\chi_{\r\r}$ vanishing as $q^2$ and as $q^4$, which may be difficult to do in practice.\footnote{Note that this situation is `softer by $q^2$' than that of the regular Bose Hubbard model, for which $\kappa\sim q^2$ in the MI, and $\kappa\sim {\rm const}$ in the `superfluid' (in quotes since there is no superflow in 1d).}
	
	An alternate diagnostic is obtained by probing correlation functions of the {\it integrated} charge density $\int^x dx' \, (\r(x') - \ob \r) = \p_x\t(x)/\twp$, which counts the density of dipolar bound states at $x$. Since $\p_x\t$ is the density of the objects that condense in the DC, it possesses power-law correlation functions in the DC and exponentially decaying correlation functions elsewhere:
	\be \left\lan \( \int_{x_1}^{x_2}dx'\, (\r(x') - \ob\r )\)^2 \right\ran \sim \begin{cases} 	\frac1{|x_1-x_2|^2} & {\rm DC}  \\  e^{-|x_1-x_2|/\xi} & {\rm else}
	\end{cases} \ee  
	giving a sharper distinction between the two phases. Quantum gas microscopes \cite{bakr2009quantum}, which can directly read off the density $\r_i$ on each site, are an ideal platform for measuring this type of correlation function.

	\section{Finite size effects and the Bose Einstein Insulator} \label{sec:bei}
	
	As we saw in Sec. \ref{sec:field_theory}, vortices in $\phi$ condense at all rational fillings, due to vortex operators $\cos(m\p_x\t)$ inevitably destabilizing the free $z=2$ fixed point \eqref{lbei} which governs the BEI. However as was discussed above, the bare strength of these vortex operators $y_m \sim e^{-m^2 c\sqrt{K_D/u}}$ can easily be extremely small. If $y_m$ is small enough, finite-size effects can cut off the RG flow at a scale where the renormalized coefficient of $\cos(m\p_x\t)$ is still small. In this case, the physics of the BEI has a chance to survive,\footnote{Since $e^{i\p_x\t}$ always has LRO, one can always Taylor expand the $\cos(\p_x\t)$ appearing in the action. Only the first terms in this expansion are relevant, and thus one could imagine tuning to a multicritical point where both $(\p_x\t)^2$ and $(\p_x\t)^4$ are absent. This gives a way of realizing BEI even in the thermodynamic limit, provided one is willing to accept the fine-tuning of two parameters. We thank Anton Kapustin and Lev Spodyneiko for this remark.} 
	and as we will see in Sec. \ref{sec:half_filling} there is some evidence for this occurring in our DMRG numerics at fractional filling. In this section we first briefly discuss some of the physical signatures of the BEI, and then show how it can in principle be stabilized in finite-sized systems by studying its emergence in a dipolar rotor model. 
	
	
	\begin{figure}
		\includegraphics[width=.37\tw]{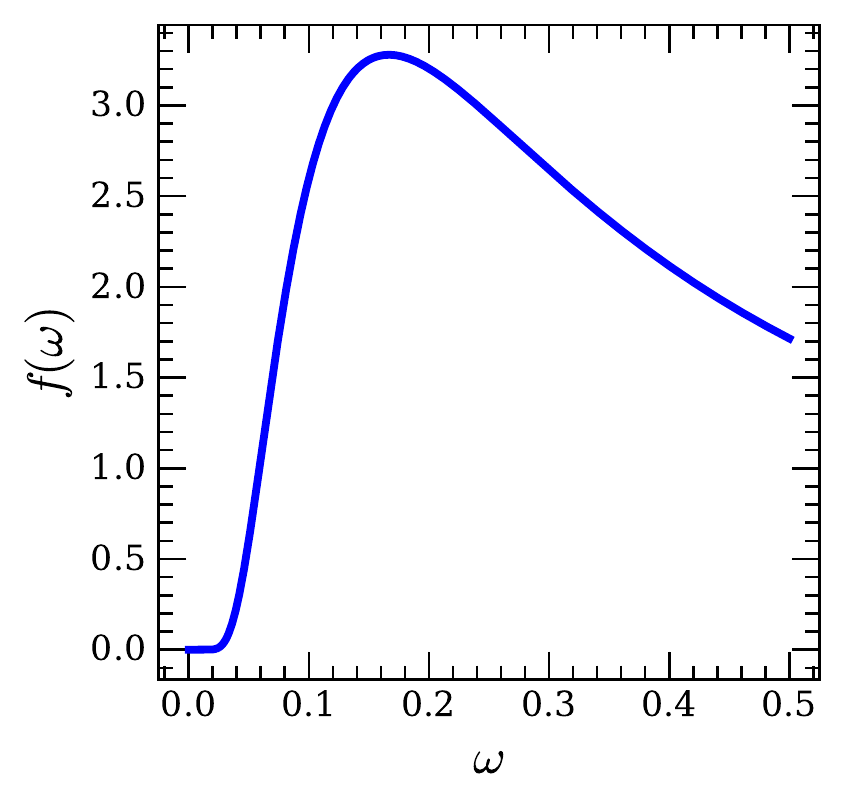}
		\caption{\label{fig:spectral_func} The function $f(\o) = e^{-1/(4\o)}/\o^{3/2}$, which is proportional to the boson spectral function $A(\o)$ in the BEI phase. Note that despite appearances, $f(\o)\neq0$ for all $\o\neq0$.} 
	\end{figure}
	
	\ss{The physics of the BEI} 
	
	When discussing the BEI, we can compute with the quantum Lifshitz model \eqref{lbei} (see Ref. \cite{gorantla2022global} for a recent discussion of various ways to interpret this continuum field theory).
	
	To determine whether the BEI has a nonzero charge gap, we first compute the boson spectral function. 
	At coincident spatial points, the boson correlator in time is evaluated as 
	\bea  \ln C_{e^{i\phi}}(\tau,0) &= \frac1{2^{3/2}( K_D  K_\tau^3)^{1/4}} \int_\o \frac{1-\cos(\o\tau)}{|\o|^{3/2}}\\ & =  \frac1{2\sqrt\pi( K_D K_\tau^3)^{1/4}} \sqrt{|\tau|}.\eea 
	so that the boson operators $b\sim e^{i\phi}$ decay exponentially in imaginary time as 
	\be C_{e^{i\phi}}(\tau,0)= e^{-\sfc \sqrt{\tau}},\qq \sfc \equiv \frac1{2\sqrt\pi ( K_D K_\tau^3)^{1/4}}.\ee 
	This tells us that the boson operators $b\sim e^{i\phi}$ have short-ranged correlation functions (in time as well, with $C_{e^{i\phi}}(0,x) \sim e^{-\sfc x}$ by the $z=2$ nature of the fixed point). However, this does {\it not} by itself imply that the bosons are gapped. 
	To determine the charge gap, we need to compute the boson spectral function $A(\o)$ by Fourier transforming, send $\o \ra -i \o + 0^+$, and take the imaginary part of the resulting expression. 
	This yields 
	\bea \label{ao} A(\o) & = -\frac1\pi {\rm Im} \[\int_\tau e^{-i\o\tau} C_{e^{i\phi}}(\tau,0) \]_{\o \ra -i\o+\ep}  \\ & \approx \frac{\sfc}{2\pi^{3/2}} \frac{e^{-\frac{\sfc^2}{4|\o|}}}{|\o|^{3/2}},\eea 
	which has an interesting essential singularity as $\o \ra 0$, with the function $f(\o) = e^{-1/(4\o)}/\o^{3/2}$ shown in Fig. \ref{fig:spectral_func}. 
	Thus while the spectral weight is suppressed dramatically at low frequencies, $A(\o) \neq 0$ for all nonzero $\o$, and strictly speaking, the bosons are gapless. 
	
	In accordance with the (barely) nonvanishing charge gap, the BEI is also checked to be compressible, with 
	\be \label{beikappa} \kappa =  K_\tau -  K_\tau^2  \(\int_{\tau,x} e^{ i q x}\lan \p_\tau\phi(\tau,x) \p_\tau\phi(0,0)\ran\)_{q\ra 0} =  K_\tau.\ee  
	{
		On the other hand, calculating the {\it equal-time} density-density correlatiors gives 
		\be \label{chibei} \c(q) \propto q^2,\ee 
		which differs from the $|q|^3$ dependence in the DC \eqref{chi_equal_time}.}
	
	Despite being a continuous symmetry, and despite being in one dimension, dipole symmetry is actually spontaneously broken in the BEI at $T=0$ \cite{lake2022dipolar,stahl2021spontaneous,kapustin2022hohenberg}: indeed, correlations of the dipole order parameter $e^{i\p_x\phi}$ go as 
	\be \label{dxphilro} C_{e^{i\p_x \phi}}(0,x) \sim e^{-\int_{q,\o}q^2 \frac{1-\cos(qx)}{\o^2 K_\tau+q^4 K_D}} \xra{x \ra \infty} 1,\ee 
	as the integral in the exponential is IR-finite (c.f. the power law behavior in the DC phase \eqref{cdphi}), implying a nonzero expectation value $|\lan d_i \ran| \neq 0$. This does not contradict the Mermin-Wagner theorem, which allows dipole symmetry to be spontaneously broken in 1d at $T=0$ \cite{kapustin2022hohenberg,stahl2021spontaneous,lake2022dipolar}, {\it provided} that the compressibility is nonzero (which in the BEI it is, according to \eqref{beikappa}).

	\ss{The BEI in a rotor model}
	
	We now demonstrate how the physics of the BEI can emerge in finite-sized systems. We will work at large integer fillings, and at hopping strengths below those set by the instability \eqref{fbd_cond}. In this regime, the DBHM can be studied by way of the rotor model 
	\be H = \frac{U}2\sum_i n_i^2 - J \sum_i \cos(\De_x^2\phi_i),\ee 
	where $[e^{i\phi_i},n_j] = \d_{i,j} e^{i\phi}$.
	Note that in this model the instability towards the FBD phase will be absent (since the microscopic degrees of freedom are rotors, rather than bosons). 
	
	This model can be easily simulated with classical Monte Carlo techniques, as we can equivalently study the 2d classical rotor model
	\be \label{hrotor} H= -\mcj \sum_i \(\cos(\De_\tau\phi) + \cos(\De_x^2\phi) \),\ee 
	where $\mcj = \sqrt{J/U}$. One advantage of doing this is that the compressibility---which is nonzero only in the BEI phase---is easy to evaluate (unlike in DMRG, where the calculation of static response functions is generally rather difficult). 
	
\begin{figure}
		\centering 
		\includegraphics[width=.4\tw]{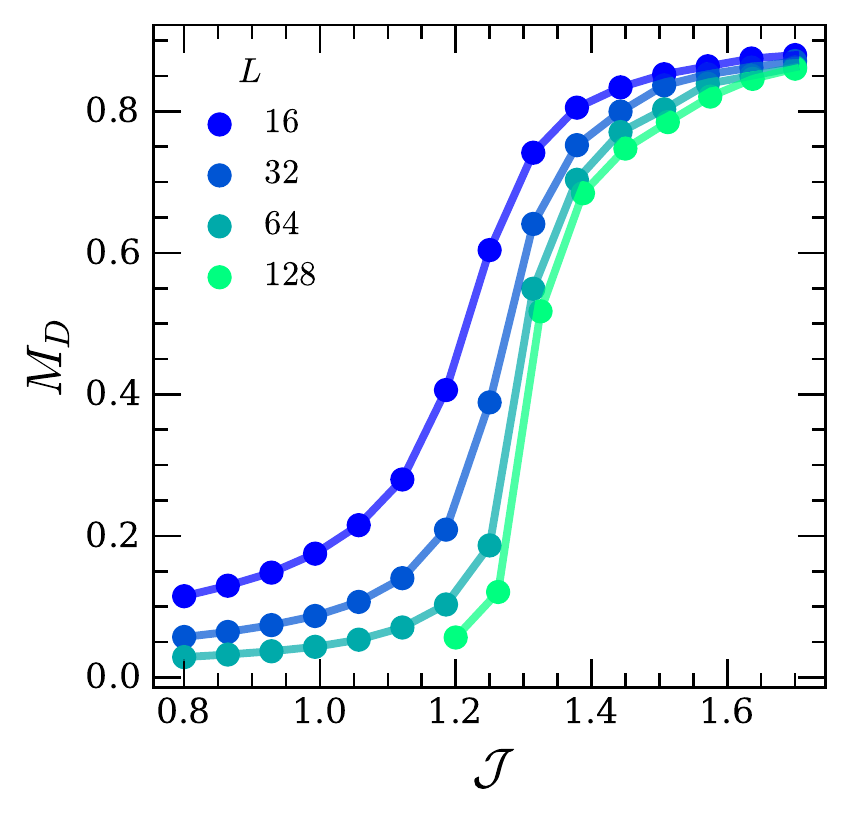} 
		\includegraphics[width=.4\tw]{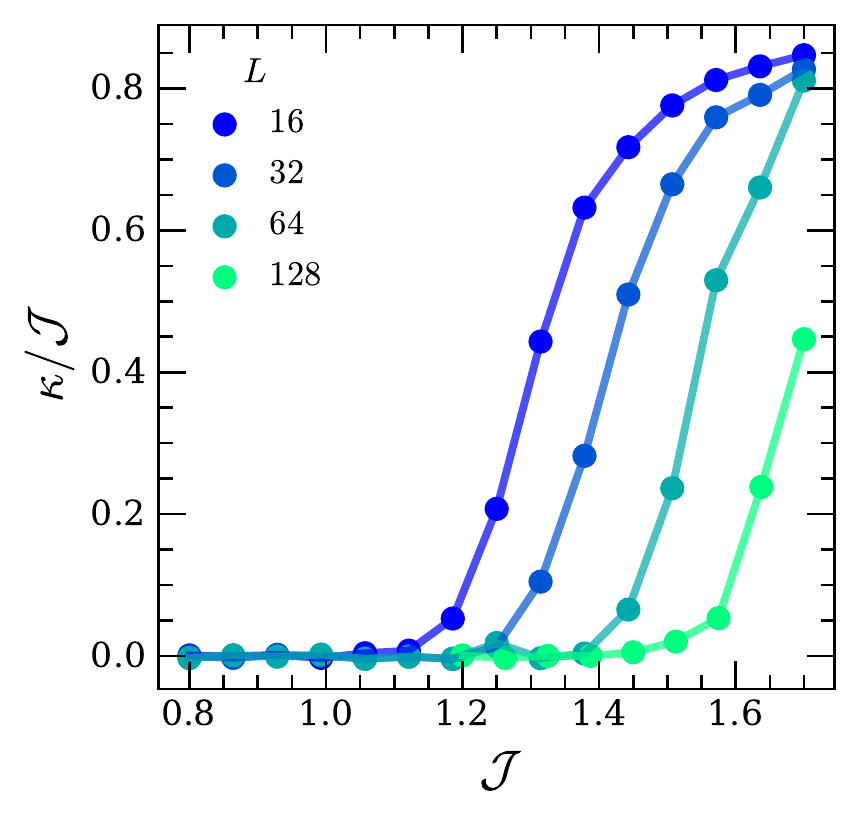} 
		\caption{\label{fig:mcfig} The dipolar magnetization (top) and compressibility (bottom) obtained from Monte Carlo simulations of the rotor model \eqref{hrotor}. } 
	\end{figure}

	Results of these simulations for square systems of linear size $L=16,\dots,128$ are shown in Fig. \ref{fig:mcfig}. In the top panel, we plot the dipolar magnetization 
	\be M_D = \frac1{L^2}\left\lan \sqrt{\(\sum_i \cos(\De_x\phi_i)\)^2 + \( \sum_i \sin(\De_x \phi_i) \)^2 } \right\ran,\ee 
	which can be used to detect the transition into the DC. We see from the plot that $M_D$ onsets at a coupling $\mcj_{c,DC}$ that converges to $\mcj_{c,DC}\approx1.25$ at large $L$. 
	
	In the bottom panel of Fig. \ref{fig:mcfig}, we plot the compressibility 
	\be \kappa = \frac{\mcj}{L^2}\sum_i\lan \cos(\De_\tau\phi) \ran - \frac{\mcj^2}{L^4} \sum_{i,j} \lan \sin(\De_\tau\phi_i)\sin(\De_\tau\phi_j)\ran,\ee
	which is zero in the MI and DC, but nonzero in the BEI. We see clearly from the plot that a nonzero compressibility onsets after some critical value $\mcj_{c,BEI}> \mcj_{c,DC}$, with the gap between $\mcj_{c,BEI}$ and $\mcj_{c,DC}$ becoming monotonically {\it larger} with increasing system size. Extrapolating this trend, we see that the BEI disappears in the thermodyanmic limit but survives at finite $L$, entirely in accord with the theoretical expectations of Sec. \ref{sec:field_theory}. 

	\section{Integer fillings: Mott insulators and dipole condensates} \label{sec:int_filling}
	
	We now turn to a slightly more detailed look at various parts of the phase diagram, starting at integer fillings $(m=1)$. 
	
	\ss{Dipolar mean field theory} 
	
	The physics at integer filling is rather simple: as the strength of the hopping terms is increased, a (continuous) transition---driven by the condensation of dipoles---occurs between the MI and the DC. The location of this transition can be identified in mean field theory by proceeding as in Ref. \cite{lake2022dipolar}. We start by writing the hopping terms in the DBHM Hamiltonian \eqref{dbhm_ham} as 
	\be H_{hop} = -\sum_{i,j} b^\da_{i}b_{i+1} [\mca]_{ij} b_{j+1}^\da b_j,\ee 
	where the matrix $\mca$ is defined as
	\be [\mca]_{ij} =t(\d_{j,i+1} + \d_{i,j+1} +\d_{i,j+2} + \d_{j,i+2}).\ee 
	To determine where the transition into the DC occurs, we decouple the hopping term in terms of dipole fields $D_i$ as 
	\be H_{hop} = -\sum_{i} \( b_i^\da b_{i+1} D_i + (D_i)^\da b^\da_{i+1} b_i\) + \sum_{i,j} (D_i)^\da [\mca]\inv_{ij} D_j.\ee 
	We then integrate out the bosons and obtain an effective action for the $D_i$, with the transition being identified with the point where the mass of the $D_i$ fields changes sign. The manipulations are straightforward and are relegated to App. \ref{app:effective_action}, where we show that the transition occurs at
	\be \label{mf} t_{DC,mf} = \frac{U}{4n(n+1)}.\ee
	The natural expectation from theory is that this transition is of BKT type, although we leave a detailed study of the critical point to future work.

	\subsection{DMRG: results and interpretation}
	
\begin{figure*}
		\includegraphics[width=0.85\tw]{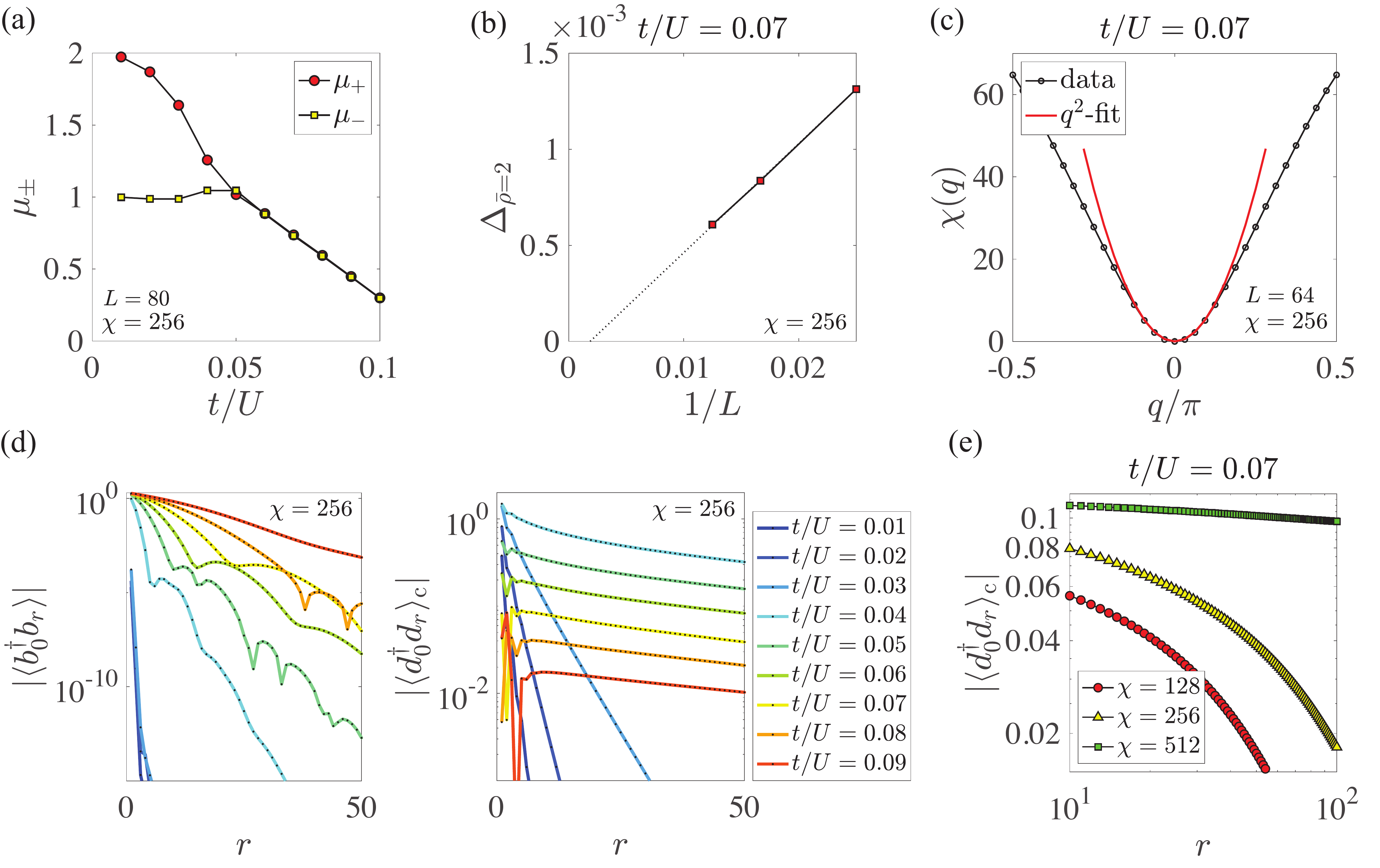}
		\caption{DMRG results at filling $\bar{\rho}=2$: (a) chemical potentials $\mu^+$ and $\mu^-$ (see text for definitions) vs. $t/U$. The asymmetry $\mu_+-\mu_-$ vanishes for $t \geq t_{DC}\approx 0.05 U$. (b) energy gap in the same boson number sector, which scales as $1/L$, indicating that the dynamical exponent $z=1$. (c) equal-time density-density correlation function vs. momentum $q$ in the DC phase. (d) the boson (left) and dipole (right) connected correlation functions at various values of $t/U$. The boson correlators decay exponentially at all $t/U$, while the dipole correlators switch to a slow power-law decay in the DC phase. (e) bond dimension dependence of the dipole-dipole connected correlator deep in the DC phase. $\c=512$ provides a good fit to a (small) power law, while for $\c=128,256$ the correlators are mean field like, and decay exponentially. Panels (a)-(c) were obtained using finite DMRG with $\chi=256$ and a small single-particle hopping of $t_0/U=10^{-4}$; periodic boundary conditions were imposed for (c). Panels (d),(e) were obtained with infinite DMRG and $t_0/U=10^{-5}$.} 
		\label{fig:rho2}
	\end{figure*}

	DMRG simulations largely conform with the above mean field picture. Before discussing the results, we briefly note that to aid in the convergence of DMRG, we have found it useful to add a small amount of dipole-violating single-particle hopping $t_0$ ($t_0/U \leq 10^{-4}$), via the term $H_{sp}$ of \eqref{hsp}. 
	From our field theory treatment we expect $H_{sp}$ to be an irrelevant perturbation throughout the phase diagram,\footnote{Except in the BEI, where $H_{sp}$ is relevant and eventually drives the system to a conventional Luttinger liquid. Despite the fact that the BEI may effectively emerge at fractional fillings due to DMRG not fully capturing the thermodynamic limit, $\lan b^\da_i b_j\ran$ is nevertheless observed to always decay exponentially for all values of $t_0$ we consider, indicating that the presence of $H_{sp}$ indeed has no effect on the IR physics.} given that the analysis of Sec. \ref{sec:expt} predicts a nonzero charge gap in every phase. This prediction is borne out in our numerics [Fig. \ref{fig:rho2} $(d)$]: the decay of $\lan b_i^\da b_j\ran$ softens with increasing $t/U$, but decays exponentially even deep in the DC. In keeping with this, the perturbation \eqref{hsp} is not observed to qualitatively change any features of the phase diagram. 
	
	The most straightforward way of identifying the DC phase is by examining connected correlation functions $\lan d_i^\da d_j\ran_c$ of the dipole operators $d_i = b_i^\da b_{i+1}$, which exhibit QLRO in the DC and are short-ranged in the MI. 
	
	First consider unit boson filling, $n= 1$. Since in our numerics we set $t_3=t_4$, the mean-field estimate \eqref{mf} of the transition from MI to DC gives $t_{DC,mf}/U = 1/8$, which interestingly matches exactly the value set by the instability of \eqref{fbd_inst}. Thus for $n = 1$ we are not guaranteed to see a DC, as mean-field theory predicts a direct transition from the MI to the FBD phase. This is indeed what occurs in DMRG, with the MI extending all the way up until the transition into the FBD phase. 
	
	While the instability that occurs when $t \geq t_{FBD}$ is independent of $n$, the mean-field prediction for the DC transition scales as $1/n^2$, and so for all $n>1$ we expect a DC to be present between the MI and FBD phases. Indeed, our numerics find that $\lan d_i^\da d_j\ran_c$ displays a sharp crossover from a rapid exponential decay to a slow power-law falloff at a critical value of $t_{DC}$, which for $n=2$ is $t_{DC} \approx 0.05U$ (see Fig. \ref{fig:rho2} $(d)$). Despite the fact that we are in 1d---where quantum fluctuations are strongest---this value agrees quite well with the mean-field prediction, which for the parameters used in Fig. \ref{fig:rho2} gives $t_{DC,mf} = U/24 \approx 0.042U$. 
	
	Deep in the DC phase, fitting the $\lan d_i^\da d_j\ran_c$ correlators to the functional form $\frac1{|i-j|^\a} e^{-|i-j|/\xi}$ gives small power-law exponents and extremely large correlation lengths. 
	Deep in the DC phase the connected correlators plotted in Fig. \ref{fig:rho2} $(d)$ ultimately fall off exponentially at large distances, and the dipole operators have a nonzero expectation value $\lan d_i\ran\neq0$. Since the DC is incompressible, only QLRO is possible in the DC (unlike in the compressible BEI; see the discussion around \eqref{dxphilro}). 
	
	The ultimate exponential decay of $\lan d_i^\da d_j\ran_c$ and the nonzero value of $\lan d_i\ran$ are thus simply due to DMRG not fully capturing the gapless fluctuations that ultimately reduce the dipole order from long-ranged to quasi-long-ranged. This is not surprising, as the suppression of LRO is logarithmically weak in the system size $L$: estimating the fluctuations in the standard way gives 
	\bea \label{suppress}
	\langle d_i \rangle &\sim \langle e^{i\p_x\phi} \rangle \sim  \lan d_i\ran_{mf} \(1-\frac{1}{2} \langle (\p_x\phi)^2 \rangle \) \\ & \sim\lan d_i\ran_{mf}\( 1- \a \log L \),
	\eea
	with $\a$ a non-universal constant determined by the correlator \eqref{cdphi}, and $\lan d_i\ran_{mf}$ the dipole expectation value in mean-field. In iDMRG, for the purposes of \eqref{suppress} we can think of the bond dimension $\c$ as producing an effectively finite $L$, and we thus expect that the LRO should be (slowly) suppressed with increasing $\c$. 
	This is indeed what we observe, with the exponential decay at $\c=256$ giving way to more-or-less pure power-law behavior by the time $\c=512$ [Fig. \ref{fig:rho2} (e)] (the very weak decay is due to being very deep in the DC phase). 
	
	{Another result of our field theory analysis is the prediction \eqref{chi_equal_time} that the static charge-charge correlator vanishes as $\c(q)\propto|q|^3$ in the DC. Fig. \ref{fig:rho2} (c) shows $\c(q)$ obtained from DMRG deep in the DC phase, which indeed vanishes polynomially with $q$. For the system size used to compute this correlator $(L=64)$, extracting the precise exponent is difficult, and a fit to $\c(q) \propto q^2$ naively appears to work better. Interestingly, $\c(q)\propto q^2$ is in fact precisely the dependence we expect in the BEI (see Sec. \ref{sec:bei}). We however do not interpret this as evidence of a BEI phase that is stabilized by finite size / finite bond dimension effects. One reason for this is that we do not see dipole correlators that convincingly have LRO, with $|\lan d_i\ran|$ very small and suppressed with increasing bond dimension. Another reason comes from our measurement of the energy gap scaling, as we now discuss. }
	
	In addition to correlation functions, we also directly measure the chemical potentials 
	\begin{align}
		\mu^+ & \equiv E_g (N+1) - E_g (N) , \nonumber \\
		\mu^- & \equiv E_g (N) - E_g (N-1) , 
	\end{align}
	obtained as the ground state energy difference of $N$, $N+1$, and $N-1$ bosons, respectively. 
	Focusing on $n=2$, plots of $\mu^\pm$ vs. $t/U$ are shown in Fig. \ref{fig:rho2} (a), where the asymmetry $\mu^+ - \mu^- $ is shown to vanish at a certain critical value which agrees well with that obtained by looking at the onset of QLRO in the dipole correlators. 
	Since 
	\begin{align}
		\mu^+ - \mu^- = E_g (N+1) + E_g (N-1) - 2E_g (N),
	\end{align}
	the fact that $\mu^+=\mu^-$ in the DC phase can be understood simply as a consequence of the DC possessing gapless particle-hole excitations. 
	
	To probe the particle-hole excitation energy more carefully, we examined the energies of the ground state and the first-excited states within the same $N$-particle sector. As shown in Fig. \ref{fig:rho2} (b), the energy difference scales as $\Delta E \sim 1/L$, consistent with the dynamical exponent $z =1$ as predicted by our field theory treatment of the DC. { Note that $z=1$ is {\it not} what is expected in the BEI, which has $z=2$; we thus take this as evidence that---at least for this filling---DMRG is able to fully account for the perturbations that render the BEI unstable in the thermodynamic limit. }
	
	Further supporting evidence is obtained by computing the entanglement entropy, which is shown in the bottom panel of Fig. \ref{fig:dmrg} (c). The dipolar nature of the Hamiltonian means that the presence of spatial boundaries have a large effect on the entanglement entropy near the chain ends, preventing a fit to the Calabrese-Cardy formula \cite{calabrese2009entanglement} from working over the entire chain length. If we however ignore the boundaries and only fit the interior $\sim$80\% of the chain, we obtain a good fit with central charge $c=1$, again matching what our field theory analysis predicts for the DC.

	\section{Half-integer filling: pair hopping models and charge density waves}  \label{sec:half_filling} 
	
	\begin{figure*}
		\includegraphics[width=0.85\tw]{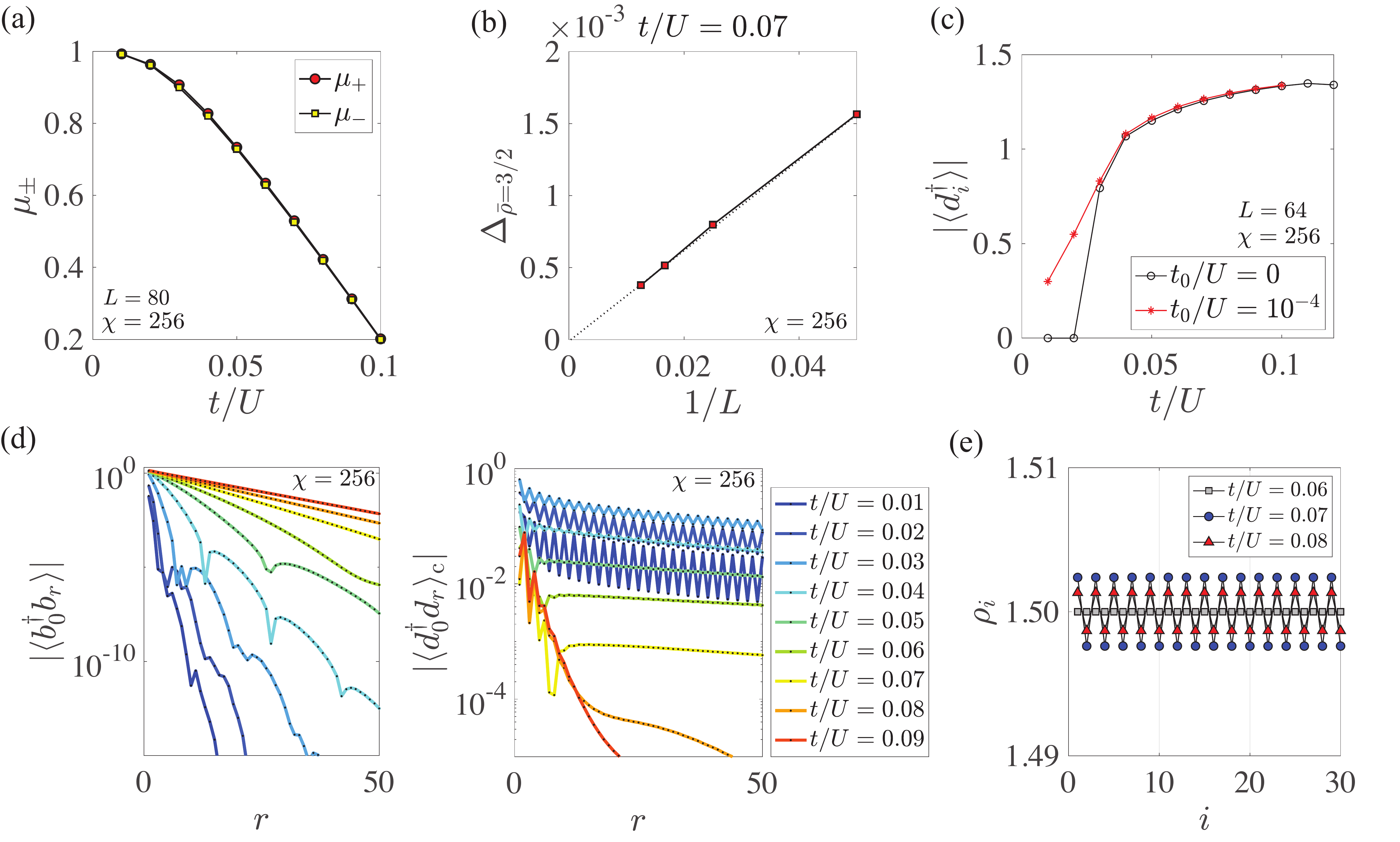}
		\caption{DMRG results at filling $\bar{\rho}=3/2$: (a) chemical potentials $\mu^+$ and $\mu^-$ vs. $t/U$. $\mu_+=\mu_-$ for all $t/U$, as expected from the particle-hole symmetry of the DC. (b) energy gap in the same boson number sector, indicating a dynamic exponent of $z=1$. (c) DC amplitude as a function of $t/U$. (d) boson (left) and dipole (right) connected dipole correlation functions, at various values of $t/U$. The boson correlators decay exponentially for all $t/U$. In the bDC phase ($t\leq t_\star \approx 0.065U$) the dipole correlators oscillate at momentum $\pi$, while the oscillations disappear at $t>t_\star$. (e) Expectation value of the boson density at different sites $i$. The grey boxes are taken at $t<t_\star$, while the other two curves are taken at $t>t_\star$, showing (weak) period-2 CDW order. 
			All DMRG hyperparameters are the same as in Fig. \ref{fig:rho2}. } 
		\label{fig:rho15}
	\end{figure*}	
	
	{
		We now come to the case of half-odd-integer fillings ($m=2$). We will see that general theoretical considerations lead to the possibility of having two distinct types of dipole condensates
		distinguished by their patterns of symmetry breaking: one spontaneously breaks site-centered reflections $R_s$ and is realized at small $t$, while the other spontaneous breaks $R_b$ and can arise at larger $t$ (the $R_s$-breaking DC exists only when $t_4$ is nonzero, and is thus unlikely to occur in the optical lattice setup). 
		
		In this section we will see how these two types of DCs can be understood within the theoretical framework developed above. Our DMRG results will be seen to confirm the existence of the $R_s$-breaking DC phase at small $t$, but for $\ob\r > 1$ and at large $t$ we seem to observe an effective BEI phase instead of the $R_b$-breaking DC. As discussed above, the BEI is presumably eventually unstable in the thermodynamic limit, but the limitations of our numerics prevent us from seeing this instability directly.  }

	We first consider what happens at the smallest values of $t$ (the regions denoted by `bDC' in Fig. \ref{fig:dmrg} (a); this terminology will be explained below). As far as the Hubbard repulsion is concerned, the lowest energy states are those with boson number $(n\pm1)/2$ on each site, and for $t/U\ll1$ we can consequently restrict our attention to the effective spin-half single-site subspace 
	\be \mch_{1/2} = \{\k\doa\equiv \k{(n-1)/2},\k\upa\equiv\k{(n+1)/2}\}.\ee 
	When restricted to $\mch_{1/2}$, the Hamiltonian reduces to\footnote{Adding a nearest-neighbor Hubbard repulsion $U'$ results in the addition of the term $\frac{U'}8 \sum_i (\s^z_i \s^z_{i+1}+4n\s^z_i)$, the presence of which leads to a period-2 CDW at the smallest values of $t_4 / U'$, which at intermediate $t_4/U'$ melts and gives way to the gapless state described below.}
	\bea \label{hhalf} H_{1/2} & = -t_4\frac{(n+1)^2}4\sum_i \s^+_i \s^-_{i+1} \s^-_{i+2} \s^+_{i+3} + h.c.,\eea
	where the $\s^\pm_i$ act on $\mch_{1/2}$. While we are still setting $t_3 =t_4 =t$, we have written $t_4$ above to emphasize that $H_{1/2}$ is trivial at leading order if $t_4=0$, since the $t_3$ hopping term has no matrix elements that act within the $\mch_{1/2}$ subspace.
	
	This spin model has appeared extensively in the literature, where it has been used to understand Krylov fracture, and---when the $\s^\pm_i$ are replaced by spinless fermion creation / annihilation operators---as a way of probing quantum Hall physics \cite{moudgalya2019thermalization,rakovszky2020statistical,moudgalya2020quantum,seidel2005incompressible}. 
	The ground state of $H_{1/2}$ can be thought of as a correlated `breathing' pattern of the state $\k{\cdots \doa\doa\upa\upa\doa\doa\upa\upa\cdots}$, i.e. a linear combination of this state and all states obtained from it under the action of all powers of $H_{1/2}$. States of this form allow the bosons room to locally resonate back and forth and thus lower their kinetic energy, while states like $\k{\cdots\upa\doa\upa\doa\cdots}$ are annihilated by $H_{1/2}$ and carry a large kinetic energy cost. We are thus prompted to define the effective spins $\k{\wt \upa}_i \equiv \k{\upa\doa}_{2i,2i+1}, \k{\wt \doa}_i \equiv \k{\doa\upa}_{2i,2i+1}$ \cite{moudgalya2019thermalization,rakovszky2020statistical}; in this representation the effective Hamiltonian is simply
	\be \label{xy} H_{1/2} = -t_4 \frac{(n+1)^2}4 \sum_i\wt\s^+_i \wt\s^-_{i+1} + h.c.,\ee 
	where the $\wt\s^\pm_i$ operate on $\wt\mch_{1/2} = \{\k{\wt \upa}, \k{\wt \doa}\}$. Thus in the limit where we can project into $\wt\mch_{1/2}$, the dipole-conserving spin-1/2 model \eqref{hhalf} can in fact simply be solved by fermionization. 
	
	As a result, the phenomenology of the small $t/U$ phase is easy to describe. For example, the dipole order parameter $d_i$ becomes $\wt\s^+_i$ if $i\in 2\zz$, while it acts outside of $\wt\mch_{1/2}$ if $i\in 2\zz+1$. This results in the correlation function of the dipole operators taking the form 
	\be \label{doscil} \lan d_i^\da d_j\ran \propto (1+\gamma (-1)^i)(1+\gamma (-1)^j) \frac1{|i-j|^\b}, \ee
	where $\b$ is a non-universal Luttinger parameter depending on $t/U$, and 
	$\gamma\leq1$ is another non-universal parameter controlling the strength of the oscillations. 
	This form for the correlator is confirmed by DMRG (performed at $\ob\r=3/2$), with both $\g$ and $\b$ decreasing with larger $t/U$ [Fig. \ref{fig:rho15} (d), right]. These oscillations can be thought of as producing a {\it bond}-centered CDW (hence the `b' in `bDC'), breaking site-centered reflections ($R_s$) but not bond-centered ones ($R_b$). 
	In contrast to dipole correlators, density correlation functions are non-oscillatory, and $\lan n_i\ran = n/2$ is observed to be uniform throughout the small $t/U$ phase [Fig. \ref{fig:dmrg} (b), panel 2], in keeping with the fact that $\lan \s^z_i\ran = 0$ in the ground state of \eqref{xy}. 
	Single bosons remain gapped in the bDC, and $\lan b_i^\da b_j\ran $ decays extremely rapidly with $|i-j|$ [Fig. \ref{fig:rho15} (d), left]. 
	The existence of a dipole condensate is further confirmed by measurements of the chemical potentials $\mu_\pm$, with $\mu_+=\mu_-$ for all $t/U$ [Fig. \ref{fig:rho15} (a)], consistent with the particle-hole symmetry of the DC. 
	
	The resonating processes described above allow the system to somewhat reduce its kinetic energy, but the motion of charges is still constrained by the projection into $\wt\mch_{1/2}$.
	As $t$ is increased, one theoretically expects that it eventually reaches a value $t_\star$ at which a phase transition into a distinct type of $R_b$-breaking DC occurs (in the region denoted simply as `DC' in the phase diagram of Fig. \ref{fig:dmrg} (a)). 
	In terms of the above spin-1/2 model defined on $\mch_{1/2}$, the existence of a transition between the two types of DC can be understood as follows.
	
	The projection from $\mch_{1/2}$ to $\wt\mch_{1/2}$ eliminated the states $\k+_i \equiv \k{\upa\upa}_{2i,2i+1}$ and $\k{-}_i\equiv \k{\doa\doa}_{2i,2i+1}$, which we now bring back. It is easy to convince oneself that neither of these states can propagate freely by themselves under the dynamics described by \eqref{hhalf}. However, the bound states $|+ \rangle_i | - \rangle_{i+1}$ or $|- \rangle_i | + \rangle_{i+1}$ {\it can} propagate \cite{moudgalya2019thermalization}, provided that they move in a background which is ferromagnetic in terms of the $\wt\upa,\wt\doa$ spins (namely all $\k{\wt\upa}$ spins in the case of the $|+-\rangle$ bound state, or all $\k{\wt\doa}$ spins in the case of $|-+\rangle$). 
	These bound states are created by the dipole operators $d_{2i+1}$ when they act on $\wt\mch_{1/2}$. This means that increasing $t$ will have the effect of promoting the formation and propagation of these bound states. 
	Further lowering of the kinetic energy is thus achieved by letting the $|\pm\mp\ran$ bound states propagate on top of a background of either $\k{\wt\upa\wt\upa\cdots}$ or $\k{\wt\doa\wt\doa\cdots}$. 
	
	When translated back into the original boson variables, the ferromagnetic states in $\wt\mch_{1/2}$ correspond to product states in which $\lan n_i\ran = (n+(-1)^i)/2$, thereby producing a period-2 site-centered CDW. This CDW differs from the bond-centered CDW at $t<t_\star$ by its pattern of symmetry breaking, breaking $R_b$ but preserving $R_s$. 
	
	{
		We close this section by taking a more detailed look at our DMRG results for $m=2$. For all $n$, our DMRG finds the $R_s$-breaking bDC phase at small $t/U$, as expected. At half-filling ($n=1$) the bDC phase is observed to extend all the way up to $t_{FBD}$, while for $n>1$ we observe a transition at a value of $t_\star < t_{FBD}$. 
		
		However, instead of transitioning into the $R_b$-breaking DC, our numerics find a transition into a BEI-like phase where the dipoles develop LRO ($|\lan d_i\ran |\neq 0$, Fig. \ref{fig:rho15} (c)). This picture is supported by the equal-time density correlator (not shown), which has a good fit to $\c(q)\propto q^2$ at small $q$ (c.f. \eqref{chibei}). As discussed at length above, seeing a BEI here is presumably due to DMRG's inability to capture the true thermodynamic limit of the flow of $\cos(2\p_x\t)$, at least barring any serendipitous fine-tuning which happens to exactly eliminate the $(\p_x\t)^2$ term from \eqref{mcl_dc}. 
		Complicating this picture slightly is the fact that the observed energy gap appears to scale as $\De \sim 1/L$ even at rather large values of $t/U$ [Fig. \ref{fig:rho15} (b)], which indicates a dynamic exponent of $z=1$, different from the BEI value of $z=2$. It thus seems possible that our numerics are simply accessing a crossover regime in which the terms that destabilize the BEI are present, but have not yet flowed to their (large) fixed-point values. 
		In any case, the difference between the putative DC regions at $\ob \r = 3/2$ and $\ob \r = 2$ is thus observed to be quite large, with the former showing fairly large signs of BEI physics and the latter appearing to be a DC throughout. Why exactly there is such a large difference between these two fillings in DMRG is currently unclear to us.
	}

	\section{Generic filling: phase separation} \label{sec:gen_filling} 
	
	\begin{figure}
		\includegraphics[width=0.5\tw]{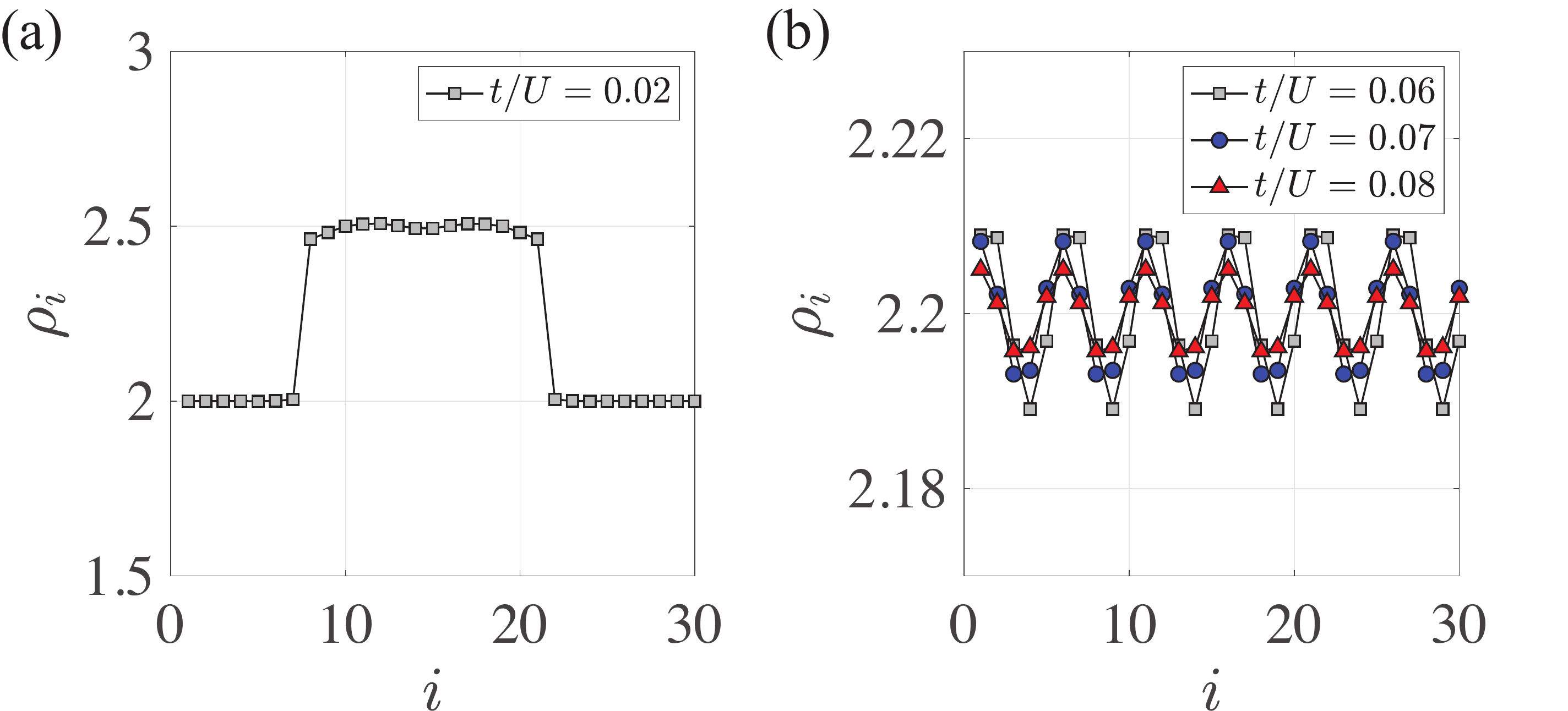}
		\caption{DMRG results at filling $\bar{\rho}=2.2$: (a) expectation value of the boson density $\rho_i$ as a function of position $i$ at small $t/U$, in the phase-separated regime. (b) $\rho_i$ in the DC, showing (weak) period-5 CDW order.  } 
		\label{fig:rho22}
	\end{figure}

	Finally, we briefly address the case of generic fillings (for the Hamiltonian \eqref{dbhm_ham}, `generic' means any $m>2$). In the absence of longer-ranged Hubbard interactions---which are not present in our simulations but appear in our field theory by way of the terms $\cos(m\t)$---the system will not be able to form a CDW in the limit of zero hopping strength. Instead, we find numerically [Figs. \ref{fig:dmrg} (b) and \ref{fig:rho22} (a)] that the system tends to phase-separate into regions of MI and regions of $R_s$-breaking condensate (although this situation may be modified in models with $t_4 = 0$). When $t/U$ is sufficiently large, the phase-separated regime is replaced by a phase possessing non-oscillatory dipole correlators $\lan d_i^\da d_j\ran$ and nonzero period-$m$ CDW order. An example of this is shown in Fig. \ref{fig:rho22} (b) which shows the boson density as a function of position at $\ob \r = 2.2 = 11/5$, displaying (weak) period-5 CDW order as expected. { We have not attempted to ascertain precisely where any BEI physics may occur in our numerics when $m>2$, since where exactly this happens is rather non-universal.}

	\section{Summary and outlook} \label{sec:disc}  
	
	In this paper we have explored the consequences of dipole moment conservation on the quantum ground states of interacting bosonic chains. Dipole conservation quenches the system's kinetic energy in a way rather distinct from the standard tricks of large magnetic fields or artificially engineered flat bands, with the quenched kinetic energy being a mix of kinetic energy and interactions. This quenching leads to several different types of exotic gapless condensates at small and intermediate hopping strengths.
	At strong hopping strengths our model develops an instability towards an unusual type of glassy ergodicity-breaking phase, which will be the subject of upcoming work \cite{fbd}. 
	
	A clear next step is to realize the DBHM in experiment. Currently, the most promising experimental platform seems to be in optical lattices, where a strong tilt potential can be created with a magnetic field gradient, enabling dipole-conserving dynamics over a long prethermal timescale. Recent studies on tilted Fermi Hubbard chains \cite{scherg2021observing,guardado2020subdiffusion} and a tilted quasi-2d boson system \cite{zahn2022formation} have focused on studying dynamical consequences of emergent dipole conservation following quantum quenches. 
	To explore the quantum ground states of these models, one needs only to prepare a Mott insulating state at large tilt and $t=0$, and then adiabatically increase the hopping strength $t$. 
	Beyond tilted optical lattices, it is also possible to directly engineer a dipole conserving Hamiltonian using bosonic quantum processors \cite{underwood2012low,wang2020efficient,ma2019dissipatively}, and it seems fruitful to investigate whether or not any other natural realizations exist. 
	
	The constraints imposed by dipole conservation have the attractive feature that they rely only on the existence of a single additional conservation law to be operative, and thus do not depend on any particular fine-tuning of the system's Hamiltonian. That said, one should not necessarily limit oneself to kinematic constraints that arise from simple conservation laws, as there are many ways in which more exotic types of kinematic constraints could be designed in principle ($e.g.$ using the Floquet driving protocols of \cite{zhao2019engineered}). For example, one could consider models of the form 
	\be H = -t \sum_i \Pi_i b_i^\da b_{i+1}  + \frac U2\sum_i n_i^2,\ee 
	where $\Pi_i$ is a projector built out of boson number operators on sites near $i$, which projects onto the subspace in which motion is possible (this is similar to e.g. the model of \cite{valencia2022kinetically}, where the constraints were placed not on boson hopping, but on boson creation / annihilation). Is there a guiding principle which helps us understand the ground state physics of models like this? 
	
	A related question is to what extent models with Hilbert space fragmentation can be studied using field theory techniques similar to those used in this work. If we enforce strict fragmentation in our model by e.g. setting sharp cutoffs $n_{max},r_{max}$ on the local Hilbert space dimension and the maximum range of the dipolar hopping terms in $H$ \cite{khemani2020localization,sala2020ergodicity}, does this necessitate any modifications to our field theory analysis? Questions of this form, along with the results of the present work, lead us to believe that it currently an opportune time for understanding the ground states of kinematically constrained many-body systems. 
	
	{
		{\it Note added:} Upon posting this work we became aware of a related study of the 1d DBHM \cite{zechmann2022}, which appeared concurrently with the present version of this paper. We are particularly grateful to the authors of \cite{zechmann2022} for correcting an important mistake in the original arxiv posting of this work, which incorrectly claimed that DMRG showed evidence for an incompressible state at $\ob \r = 3/2, t>t_\star$. }
	
	\section*{Acknowledgments} 
	
	We thank Ehud Altman, Soonwon Choi, Johannes Feldmeier, Byungmin Kang, and Alexey Khudorozhkov for discussions, and Mike Hermele and Quan Manh for collaborations on related work. EL was supported by the Hertz Fellowship. TS was supported by US Department of Energy grant DE-SC0008739, and partially through a Simons Investigator Award from the Simons Foundation. This work was also partly supported by the Simons Collaboration on Ultra-Quantum Matter, which is a grant from the Simons Foundation (651446, TS). H.-Y.L. was supported by National Research Foundation of Korea under the grant numbers NRF-2020R1I1A3074769 and NRF-2014R1A6A1030732. JHH was supported by  NRF-2019R1A6A1A10073079. He also acknowledges financial support from EPIQS Moore theory centers at MIT and Harvard.
	
	\bibliography{1d_dbhm}
	
	\begin{widetext}
		
		\appendix 
		
		\section{Lattice duality} \label{app:duality}

		In this appendix we will use a slightly modified version of standard particle-vortex duality (see e.g. Ref. \cite{savit1980duality} for a review) to derive a field theory that can be used to understand the phase diagram of the 1d DBHM. The manipulations to follow are quite similar to the ones performed when dualizing a classical 2d smectic \cite{zhai2021fractonic}, the theory of which shares many parallels with the present dipole conservering model.  
		
		Our starting point is the imaginary-time lattice model 
		\be \label{phimcl} \mcl = in \p_\tau\phi + \frac{(\twp)^2u}2(n-\ob\r)^2  - K_D \cos(\De_x^2 \phi),\ee 
		where $\phi\approx \phi+\twp$ is a compact scalar field identified with the phase mode of the $b$ bosons as in \eqref{brep}, $n$ is an operator conjugate to $e^{i\phi}$ that parametrizes density fluctuations (not to be confused with the $n$ in $\ob \r = n/m$), and the definitions of the couplings $u,K_D$ are as in \eqref{kd_and_v} (we will restrict our attention throughout to the case where $u>0$). 
		This lattice model arises from taking the rotor limit of $H_{DBHM}$, which strictly speaking is valid only at large average fillings (since the $n$ appearing in \eqref{phimcl} has eigenvalues valued in $\zz$, rather than in $\nn$). Nevertheless, the rotor limit suffices to understand the much of the qualitative physics of the regular Bose-Hubbard model at all densities, and we will see that in the present context it does a similarly good job at explaining the phase diagram. 
		
		If we could Taylor expand the cosines in \eqref{phimcl}, we would obtain a quantum Lifshitz model, which is the field theory of the Bose-Einstein insulator phase described in Ref. \cite{lake2022dipolar} and investigated in detail in Ref. \cite{gorantla2022global}. However, the legitimacy of such an expansion rests on the assumption that vortices in $\phi$ can be ignored,\footnote{Unlike in the setting of e.g. Refs. \cite{gorantla2021modified,gorantla2022global},  it is not appropriate for us to work with a model that excludes vortices by hand.} and as we will see in the following, this is actually {\it never} the case in the thermodynamic limit. 
		
		To understand the effects of vortices we switch to a 2+0d spacetime lattice and Villainize the above Lagrangian, giving 
		\be \mcl =  i \ob\rho (\De_\tau \phi-m_\tau-A_\tau) + \frac{1}{8\pi^2u}(\De_\tau\phi-m_\tau-A_\tau)^2 + \frac{K_D}2(\De_x^2\phi-m_x-\De_x A_x)^2 + i h\phi,\ee 
		where the $m_\tau,m_x \in \twp \zz$ are path-summed over, and we have added the background field $A_\mu = (A_\tau,A_x)$ as well as the source field $h= \sum_i q_i\d(x-x_i) \d(\tau-\tau_i), \, q_i\in \zz$, which will be used to calculate correlation functions. $m_\tau$ lives on the temporal links of the lattice, while $m_x,$ and $h$ live on the sites. If desired we could also couple to a background gauge field $A^D_\mu$ for the $U(1)_D$ dipole symmetry. However, $A^D_\tau$ is rather ill-defined (as only the total dipole charge, rather than local dipole density, is well-defined), while $A^D_x$ enters in the same way as does $\De_x A_x$, and therefore is redundant.   
		
		We then integrate in a $\rr$-valued vector field $J = (J_\tau, J_x)$ which lives on the links of the lattice:
		\be \mcl = \frac{4\pi^2 u J_\tau^2}{2} + \frac{J_x^2}{2K_D} + i(J_\tau+\ob\r)(\De_\tau \phi - m_\tau-A_\tau) + iJ_x (\De_x^2 \phi - m_x - \De_xA_x) + i h\phi ,\ee
		where we have chosen to write the temporal part of $J$ as $J_\tau + \ob \r$ for later convenience. 
		
		Integrating out $\phi$ tells us that 
		\be \De_\tau J_\tau - \De_x^2 J_x = h \quad \implies\quad   J_\tau = \frac1\twp\( \De_x^2 \t - \frac{\De_\tau }{\D^2} h\),\quad J_x = \frac1\twp\( \De_\tau \t + \D^{-2} h\),\ee 
		where $\t$ is defined on the temporal links, and we have let $\D^2 \equiv -\De_\tau^2 - \De_x^2$ denote the lattice Laplacian. We then substitute this expression for $J_\mu$ into the above Lagrangian, and recognize that the terms which mix $h$ and $m_\tau,m_x$ can be ignored, on the grounds that they are linear combinations of delta functions with weights valued in $i\twp\zz$. Therefore we may write 
		\bea \mcl & = \frac u{2} (\De_x^2\t - \De_\tau \D^{-2} h)^2 + \frac1{8\pi^2K_D} (\De_\tau \t + \D^{-2} h)^2 - i \frac{\t}\twp (\De_x^2 m_\tau - \De_\tau m_x) \\ 
		& \qq -i\frac{A_\tau}\twp (\De_x^2 \t - \De_\tau \D^{-2} h) - i\frac{\De_x A_x}\twp (\De_\tau \t + \D^{-2} h) - i\ob\rho(m_\tau+A_\tau).\eea
		
		From the coupling to $A_\tau$, we see that the density is represented in this approach as 
		\be \label{density} \r = \ob \r + \frac1\twp\De_x^2 \t,\ee
		agreeing with \eqref{brep} in the main text. 
		Note that as $\phi$ is dimensionless, $[u] = [\De_\tau / \De_x]$ and $[K_D] = [\De_\tau / \De_x^3]$, implying that $[\t] = [1/\De_x]$, consistent with the above expression for $\r$. 
		
		\eqref{density} implies that an infinitesimal spatial translation by an amount $\mu(x)$ acts on $\t$ as 
		\be  \label{thetatform} T_\mu : \t(x) \mt (1-\De_x\mu) \t(x+\mu) + \twp \ob \r \int_{-\infty}^x dx'\, \mu(x')+ \cdots,\ee
		where the $\cdots$ are terms higher order in $\mu$ and its derivatives. 
		Formally, \eqref{thetatform} can be derived by requiring that $\r(x)$ transform as a density under a {\it spatially-varying} translation through $\mu(x)$, viz. by requiring that 
		\be T_\mu : \r(x) \mt (1+\De_x\mu)\r(x+\mu)\ee 
		to linear order in $\De_x\mu$ and derivatives thereof. Indeed, dropping higher derivatives of $\mu$, we see that under \eqref{thetatform}, 
		\bea \r & \mt \ob \r + \frac1\twp \De_x^2  \( (1-\De_x\mu) \t(x+\mu) + \twp \ob \r \int^x_{-\infty} dx' \, \mu(x') \) \\ 
		& = \ob \r(1 + \De_x \mu)  +  \frac1\twp (1-\De_x\mu)\De_x^2\t(x+\mu) \\
		& = \ob \r (1+\De_x \mu) + \frac1\twp (1-\De_x\mu)(1+\De_x\mu)^2 \De_{x+\mu}^2 \t(x+\mu) \\ 
		& = (1+\De_x\mu)\(\ob \r + \frac1\twp \De_{x+\mu}^2\t(x+\mu) \) \\ & = (1+\De_x\mu)\r(x+\mu)\eea 
		as required. 
		In particular, for uniform translations $\De_x\mu = 0$, we have 
		\be T_\mu : \t(x) \mt \t(x+\mu) +\twp \ob \r x\mu ,\qq \De_x \t(x) \mt \De_x\t(x+\mu) + \twp\ob \r.\ee 
		
		We will find it helpful to define the field 
		\be\label{ctdef} \ct = \t + \pi\ob \r x^2,\ee 
		which satisfies $\De_x^2 \ct =\twp \r$ and which is invariant under infinitesimal translations to linear order (the order we have given the action of $T_\mu$ to), in that $T_\mu : \ct(x) \mt \ct(x+\mu) + O(\mu^2)$. We may thus write the part of $\mcl$ involving $m_\tau,m_x$ as 
		\be \mcl \supset -i \sfm \ct,\qq \sfm \equiv  \frac{\De_x^2 m_\tau - \De_\tau m_x}\twp.\ee 
		Now the object $\sfm$ is an integer satisfying $\int \sfm = \int x \sfm = 0$, where $\int$ implicitly means a discrete sum over spacetime lattice points. In the usual approach to particle-vortex duality one would only have the constraint $\int \sfm = 0$ (net zero vortex number); here the extra constraint $\int x \sfm$ has the effect of enforcing zero dipole moment of the objects created by $e^{i\t}$ (which turn out to be vortices of $\De_x\phi$). However --- as in the standard case --- the physically correct thing to do is to simply ignore the topological constraint on the sum over $\sfm$, and to then use cosines of $\t, \De_x\t$ to softly enforce the delta function constraints implemented by the sum over $\sfm$. 
		
		In more detail, the cosine terms are generated as follows. Until now, all of our manipulations have been exact, and we have remained on the lattice. In order to obtain a useful EFT, we need to integrate out short-distance degrees of freedom and produce an effective action for slowly-varying fields, giving a theory with a suitable continuum limit. To do this, from the sum over $\sfm$ we select out those configurations which involve products of terms involving products of a small number of $e^{i\ct}$ operators. For us the important operators are $e^{i\ct}$ itself and $e^{i\De_x \ct}$; other operators are either already taken into account by the free part of the action (e.g. $e^{i\De_x^2 \ct}$) or else will end up being irrelevant in the final continuum theory (e.g. $e^{i \De_\tau^2 \ct}$).  
		Keeping only the configurations of $\sfm$ that generate these terms, the partition function is 									\be \label{fullz} \mcz = \prod_{q,r=0}^\infty \sum_{n_q,n_r=0}^\infty  \frac{2^{n_q + n_r}}{n_q! n_r!} \left\langle\int \prod_{j,k=1}^{n_q,n_r} dx_j \, d\tau_j\, dx_k \, d\tau_k \cos(q \ct(\tau_j,x_j)) \cos(r \De_x \ct(\tau_k,x_k))\right\rangle,\ee
		where the expectation value is with respect to the free (quadratic) part of the lattice action for $\t$.
		
		The $\t$ fields implicitly (via \eqref{ctdef}) appearing in the above expression for $\mcz$ are not the variables we aim to write our EFT in terms of, as they are defined on the lattice and contain fluctuations at short scales. To obtain a field theory, we decompose $\t = \t_s + \t_f$ into slow and fast components, where the division between `slow' and `fast' occurs at a short-distance cutoff of $1/\L\gg a$ in space (we do not impose any cutoff in frequency, partly for convenience and partly because the important distinctions between the various cosines we will generate will be spatial). 
		
		We will regulate the products of cosines appearing in \eqref{fullz} by tiling the spacetime lattice into patches of linear size $\L\inv$, requiring that no two operator insertions appear within a distance of $\L\inv$ from one another. 
		The correlation functions of $\t_f$ are local in spacetime, falling off in $\tau$ over the timescale $\L^{-2} / \sqrt{u K_D}$, and falling off in $x$ over $\L\inv$. 
		For the purposes of this discussion, it is sufficient to approximate this behavior as giving 
		\be \lan e^{iq\t_f(\tau,x)/a} e^{-iq\t_f(\tau',x')/a}\ran \sim \begin{dcases} e^{-q^2\twp \frac{\sqrt{K_D/u }}{\L a^2}} \quad & \text{$(\tau,x)$ and $(\tau',x')$ in same patch} \\ 
			0 \quad & \text{else} \end{dcases} ,\ee 
		where the factor in the exponential comes from doing the integral $\int_\rr d\o\, \int_{\L}^{a\inv}dq\,  (q^4/K_\tau + \o^2/K_D)\inv$, and where we have momentarily restored the lattice spacing $a$. This exponential factor defines the dipole vortex fugacity (this terminology will become clear shortly)
		\be y_D \equiv 2e^{-c_D\sqrt{K_D/u}},\ee 
		where the non-universal constant $c_D = \pi/( \L a^2)$ in the present crude model. Correlators of $e^{ir \De_x \t_f}$ give a similar result, but with $y_D$ replaced by the vortex fugacity $y$, defined as 
		\be y \equiv 2e^{-c\sqrt{K_D/u}},\ee 
		with $c = \pi\L $ in the present model. 
		
		Performing the integral over $\t_f$ in \eqref{fullz} simply adds factors of $(y_D^{q^2})^{n_q} (y^{q'^2})^{n_{q'}}$ and replaces occurances of $\t$ with $\t_s$ (which we consequently re-label as $\t$). The last thing to do is to recognize that while $\t$ is now (by construction) a slowly-varying field (i.e. slowly varying on the scale of the lattice spacing), $\ct$ is not if $\ob \r \neq 0$. Cosines $\cos(q \ct),\cos(r\De_x \ct)$ thus oscillate rapidly on the lattice scale and can be dropped, {\it unless} $q,r\ob \r \in \nn$, in which case $\cos(q\ct) = \cos(q\t), \cos(r\De_x\ct) = \cos(r\De_x\t)$.\footnote{From \eqref{ctdef} one might think that $\cos(m\ct)$ would be translation invariant only if $ n \ob \r \in 4\pi \nn$, but this is only because we have not been writing the $O(\mu^2)$ piece of the transformation of $\t$ under $T_\mu$.} 						Let us write the average density as 
		\be \ob \r = n/m,\qq m,n \in \nn,\quad \gcd(m,n) =1.\ee  
		The cosines in \eqref{fullz} can be re-exponentiated, and after we drop those which vary rapidly on the lattice scale, we obtain the effective continuum Lagrangian 
		\bea \label{full_mcl} \mcl & =  \frac u{2} (\De_x^2\t - \De_\tau \D^{-2} h)^2 + \frac1{8\pi^2 K_D} (\De_\tau \t + \D^{-2} h)^2 -i\frac{A_\tau}\twp (\De_x^2 \t - \De_\tau \D^{-2} h) - i\frac{\De_x A_x}\twp (\De_\tau \t + \D^{-2} h) \\ & - \sum_{q \in m\nn}\( y_{D,4q}\cos(4q\t) + y_q \cos(q\De_x\t)\),\eea
		where $y_q\propto y^{q^2}$ and $y_{D,q}\propto y_D^{q^2}$ (and the factor of $4$ in $\cos(4q\t)$ is due to the action of $R_b$ reflection symmetry \eqref{refl}). 
		After dropping the background fields, this agrees with the Lagrangian \eqref{mcl} quoted in the main text (after integrating out $\phi$ in the later). 
		
		At {\it any} rational filling, the cosines of $\De_x\t$ destabilize the $z=2$ free fixed point of the quantum Lifshitz model that one arrives at upon Taylor expanding the cosines in \eqref{phimcl}. Indeed, it is easy to check that at this fixed point $e^{iq\De_x \t}$ has long range order for all $q$, and hence the leading nonlinearity $\cos(m \De_x\t)$ will always be relevant,\footnote{We focus solely on $\cos(m\De_x\t)$ not because it is more relevant than $\cos(lm\De_x\t)$ for integer $l>1$, but because the bare coefficients of these terms are expected to be exponentially suppressed with $l$.} giving a nonzero expectation value to $\De_x\t$. Note that as $\De_x\t$ is charged under translation, translation will generically be spontaneously broken, with the system having some kind of CDW order at {\it all} non-integer rational fillings. 
		
		After expanding $\cos(m\De_x\t)$, we obtain 
		\bea \label{dc_mcl} \mcl & = \frac{m^2y_m}2 (\De_x\t - \lan \De_x \t\ran)^2  + \frac u{2} (\De_x^2\t - \De_\tau \D^{-2} h)^2 + \frac1{8\pi^2 K_D} (\De_\tau \t + \D^{-2} h)^2 + i\frac{A_x}\twp \De_x \De_\tau \t  - i\frac{A_\tau}\twp \De_x^2\t -  y_{D,4m}\cos(4m\t).\eea
		where we have dropped the unimportant coupling between $A_\mu$ and $h$ and kept only the leading cosine of $\t$, whose scaling dimension is
		\be \label{dcos} \De_{\cos(4m\t)} =  8m^2 \sqrt{\frac{ K_D}{y_{D,m}}}.\ee 
		When this cosine is irrelevant, we obtain a free $z=1$ compact scalar, which describes the DC. When it is relevant the DC is destroyed, leading to a Mott insulator at integer filling, or a translation-breaking state with gapped dipoles at non-integer rational filling. However, since $y_{D,m}$ is exponentially small in $m^2 \sqrt{K_D/u}$, the scaling dimension \eqref{dcos} can be made extremely large (particularly at nearly incommensurate fillings and large densitites [as $K_D \propto \ob\r^2$]), thus in principle leading to a DC which extends down nearly to $t = 0$. Similarly, although we have concluded that in the thermodynamic limit this system is always incompressible, the flow away from the free $z=2$ theory (which is compressible) can be very weak, due to the smallness of $y_{D,m}$. Indeed, the results of Sec. \ref{sec:bei} give a numerical study indicating that finite-size effects can be strong enough to prevent the $\cos(m\De_x\t)$ term from growing to the point where it dominates the physics, leaving a range of parameters where the system is effectively compressible, and describable by the quantum Lifshitz model. 
		
		Finally, we use \eqref{dc_mcl} to compute correlation functions of exponentials of $\phi$ (and derivatives thereof) in the DC. Setting $A_\mu = 0$ and integrating out $\t$, the free energy as a function of the source $h$ is seen to be
		\bea \ln  Z[h] & = - \frac12 \int_{q,\o} |h_{q,\o}|^2 \frac1{(\o^2+q^2)^2} \( \frac{\o^2}{\ob K_\tau} + \frac1{\ob K_D} -\(\frac{q^2}{\ob K_\tau} + \frac1{\ob K_D}\)^2 \frac{\o^2}{\o^2/\ob K_D + q^2 m^2 y_m + q^4/\ob K_\tau} \), \eea
		where $\ob K_\tau \equiv 4\pi^2 / u, \ob K_D \equiv 4\pi^2 K_D$. 
		Since we are only interested in the IR behavior of the correlators in question, we can drop the $q^2/\ob K_\tau$, $q^4/\ob K_\tau$, and $\o^2/\ob K_\tau$ terms; this then gives us the result quoted in \eqref{phiphi_main}.

		\section{Dipolar hopping from a strongly tilted potential} \label{app:hopping} 
		
		Consider bosons hopping on a 1d lattice tilted by a strong potential $V$:
		\be H = \sum_i \( -t_0( b_i^\da b_{i+1} + b_{i+1}^\da b_i) - \mu n_i + \frac U2 n_i(n_i-1) + \frac{U'}2 n_i n_{i+1} + V i n_i \) \equiv H_t + H_U + H_V,\ee
		where $H_U$ includes the chemical potential and both the onsite $U$ and nearest-neighbor $U'$ Hubbard interactions. While the bare value of $U'$ will essentially always be negligible in optical lattice setups, we include a nonzero $U'$ in the subsequent calculations for the purposes of illuminating the structure of the terms produced by the perturbation theory expansion, and because a sizable $U'$ could very well be present in other physical realizations outside of the optical lattice context. 
		
		In the limit $V \gg t_0,U$, this theory has emergent dipole conservation over a prethermal timescale which is exponentially large in $V/t$ \cite{khemani2020localization}. Our goal is to perform a rotation into a basis in which the Hamiltonian commutes with the dipole chemical potential $V \sum_i i n_i$ up to some fixed order in $t_0/V, U/V$, and to derive the strength of the resulting dipole-hopping terms. This calculation has already been performed for the closely related fermionic models of \cite{scherg2021observing,moudgalya2019thermalization}; below we simply perform the generalization of these calculations to the present bosonic model. 
		
		As in \cite{scherg2021observing,moudgalya2019thermalization}, we use a Schrieffer-Wolf transformation to rotate the Hamiltonian into a basis where it commutes with the dipole term $H_V$, working perturbatively in $t_0/V, U/V$. We write the transformed Hamiltonian as 
		\be e^\L H e^{-\L} = \sum_{k \geq 0} \frac1{k!} {\rm Ad}_\L^k(H),\ee 
		where ${\rm Ad}_\L(\cdot) = [\L ,\cdot]$ and $\L$ is anti-Hermitian.
		
		Note that it is already clear that interactions are required for producing a nonzero dipolar hopping term. Indeed, without the interaction term, $H$ is built solely of 2-body terms --- we can thus choose $\L$ to be a 2-body operator, and ${\rm Ad}_\L^k(H)$ will consequently always itself be built from 2-body operators, which can only either be purely onsite or dipole non-conserving. In fact if we just take 
		\be \L = \L_t \equiv  \frac {t_0}V\sum_i (b^\da_i b_{i+1} - b_{i+1}^\da b_i),\ee 
		it is easy to check that when $U=U'=0$,
		\be [\L_t ,H]  = [\L_t ,H_V] = t_0\sum_i (b_i^\da b_{i+1} + b_{i+1}^\da b_i) = -H_t.\ee 
		Since this is just the negative of the hopping term, the first order part ${\rm Ad}_\L(H_V)$ dutifully kills $H_t$. Moreover, since $[\L_t,[\L_t,H_t+H_V]] = 0$, the effective Hamiltonian stops at linear order, and we simply obtain $H_{eff} = e^{\L_t}( H_t + H_V) e^{-\L_t} = H_V$, which is purely onsite. This means that when $U=U'=0$, no effective dipole hopping terms are generated --- there is perfect destructive interference between all putative hopping processes, and no such processes are generated to all orders in perturbation theory. 
		
		Let us then bring back the interactions. We take 
		\be \L =\sum_{n=1}^\infty \L_n,\ee 
		where $\L_n$ is order $n$ in $t_0/V, U/V,U'/V$, and we set $\L_1 = \L_t$. 
		We fix the second order term $\L_2$ by requiring that it cancel the off-diagonal (with respect to dipole charge) terms generated when commuting $\L_1 = \L_t$ against $H_U$. Specifically, we require 
		\be  \label{l2hv} [\L_2,H_V] = -(1-\mcp) [\L_t, H_t + H_U](1-\mcp),\ee
		where $1-\mcp$ projects onto the off-diagonal component. 
		Keeping terms to third order in this expansion, $H_{eff}$ becomes \cite{moudgalya2019thermalization}
		\be H_{eff} = H_V + H_U + [\L_2 ,H_U] + \frac12 [\L_2 - \L_t,H_t] + [\L_3,H_V] + \frac13[\L_t,[\L_t,H_t]] + \frac12 \mcp [\L_t,H_t] \mcp. \ee 
		We then need the commutators $[\L_t,H_t], [\L_t,H_U]$, the evaluation of which is straightforward. Define the hopping operators 
		\be T^\pm_{i,j} \equiv b^\da_i b_j \pm b^\da_j b_i,\ee
		which among other identities satisfy 
		\bea \label{tpmid} \sum_i [T_{i,i+1}^\pm,n_j] & = T^\mp_{j-1,j} - T^\mp_{j,j+1}  \\  [T^\pm_{i,i+1},H_V] & = VT^\mp_{i,i+1} \\ \sum_j [T^s_{i,i+1},T^{s'}_{j,j+1}] & = T^{-ss'}_{i,i+2} - T^{-ss'}_{i-1,i+1} + (s-s')(n_{i+1}-n_i).\eea 
		Then 
		\bea \label{comms} [\L_t,H_t] & = \frac{t^2}V \sum_i \( b_{i+2}^\da b_{i} + 2n_{i+1} + b_i^\da b_{i+2} - (i \ra i+1)\) = 0, \\
		[\L_t, H_U] & = \frac{tU}{2V} \sum_i \{ n_i, T_{i-1,i}^+- T_{i,i+1}^+\} + \frac{tU'}{2V} \sum_i \( (T^+_{i-1,i} - T^+_{i,i+1})n_{i+1} + n_i (T^+_{i,i+1} - T^+_{i+1,i+2}) \) 
		\eea 
		\\			\\						\\			\\\\			\\						Note that $[\L_t , H_U]$ is purely off-diagonal, so the insertions of $1-\mcp$ in \eqref{l2hv} have no effect and can be ignored. 
		
		We now determine $\L_2$ via \eqref{l2hv}, which by virtue of the above now reads 
		\be [\L_2,H_V] =- [\L_t,H_U].\ee 
		This can be done in a rather brute force way by expanding $\L_2$ as a general linear combination of all 4-boson operators which are allowed to contribute, but it is simpler to simply use the middle identity in \eqref{tpmid} as inspiration, noting that one only need flip the $T^+$s to $T^-$s in \eqref{comms} to make everything work out:
		\be \L_2 = -\frac{t_0U}{2V^2} \sum_i \{ n_i, T_{i-1,i}^-- T_{i,i+1}^-\} - \frac{t_0U'}{2V^2} \sum_i \( (T^-_{i-1,i} - T^-_{i,i+1})n_{i+1} + n_i (T^-_{i,i+1} - T^-_{i+1,i+2}) \),\ee 
		which is purely off-diagonal.

		The effective Hamiltonian to cubic order is then 
		\be H_{eff} = H_V + H_U + [\L_2,H_U] + \frac12[\L_2,H_t] + [\L_3,H_V].\ee 
		$\L_3$ is chosen to kill the off-diagonal part of $[\L_2,H_U] + \frac12[\L_2,H_t]$.\footnote{This is not just an arbitrary choice: $\L_3$ cannot be chosen to cancel any of the {\it diagonal} terms, as one can show that if $[\mco,H_V] \neq 0$, then $[H_V,[\mco,H_V]] \neq 0$ for any boson operator $\mco$ --- thus $[\L_3,H_V]$ must necessarily be off-diagonal.} It is easy to see that $[\L_2,H_U]$ is purely off-diagonal, while $[\L_2,H_t]$ can have diagonal components, as both $\L_2,H_t$ are off-diagonal. Therefore the diagonal part $\mcp [\L_2, H_t]\mcp$ survives in $H_{eff}$, which to cubic order is consequently 
		\be H_{eff} = H_V + H_U + \frac12 \mcp [\L_2,H_t]\mcp.\ee 
		
		All that remains is therefore the calculation of $\mcp [\L_2,H_t] \mcp$. This is 
		\bea  \frac12 \mcp [\L_2,H_t]\mcp &   = \frac{t_0^2}{4V^2} \sum_i \mcp \Big[ U \( \{ n_i,[T^-_{i-1,i} -T^-_{i,i+1},T^+_{j,j+1}]\} + \{ T^-_{i-1,i} - T^-_{i,i+1},[n_i,T^+_{j,j+1}]\} \) \\ 
		& + U' \Big( (T^-_{i-1,i} - T^-_{i,i+1})[n_{i+1},T^+_{j,j+1}] + [T^-_{i-1,i} -T^-_{i,i+1} , T^+_{j,j+1}] n_{i+1} + n_i [T^-_{i,i+1} - T^-_{i+1,i+2},T^+_{j,j+1}] \\ & + [n_i , T^+_{j,j+1}](T^-_{i,i+1} - T^-_{i+1,i+2}) \Big) \Big] \mcp, \eea 
		which we evaluate using \eqref{tpmid} together with  
		\be  \mcp T^-_{i-1,i} T^-_{i,i+1} \mcp = - (b^\da_{i-1} b^2_i b^\da_{i+1}  + h.c.),\qq \mcp(T^-_{i-1,i} T^-_{i+1,i+2} )\mcp = - (b_{i-1} b^\da_i b^\da_{i+1} b_{i+2} + h.c.)\ee
		to write  
		\bea   \frac12 \mcp [\L_2,H_t]\mcp &   = \frac{t_0^2}{V^2} \sum_i \big(-2(U-U'/2)n_i^2 + 2(U-U') n_i n_{i+1} + U' n_i n_{i+2} \\ 
		& \qq - U' b_{i-1}^\da b_i b_{i+1} b_{i+2}^\da  -(U-U')b_{i-1}^\da b_i^2 b_{i+1}^\da + h.c.\big) \eea 
		and so the effective dipolar Hamiltonian to cubic order in $t_0/V, U/V$ is 
		\bea H_{DBHM} &=- \sum_i \( \frac{t_0^2 (U-U')}{V^2}  b_i^\da b_{i+1}^2 b_{i+2}^\da  + \frac{t_0^2 U'}{V^2} b^\da_{i-1}b_ib_{i+1}b^\da_{i+2} + h.c.\) + \sum_i (-(\mu+U/2) + i V) n_i  \\ 
		& \qq +\sum_i \( \(\frac U2 - \frac{2t_0^2(U-U'/2)}{V^2}\)  n_i^2 + \(\frac{U'}2 + \frac{2t_0^2 (U-U')}{V^2}\) n_in_{i+1}  + \frac{t_0^2U'}{V^2} n_i n_{i+2} \).\eea
		Note in particular that the effective 3-site hopping term has strength $t_3$ proportional to the {\it difference} of the onsite and nearest-neighbor interaction strengths, while the strength $t_4$ of the 4-site hopping term is proportional to $U'$, and is thus only present when the microscopic model has nearest-neighbor repulsive interactions.\footnote{At least to the present order. Since an effective nearest-neighbor repulsion is generated at third order even at $U'=0$, an effective $t_4$ term will always be generated at sixth order.}

		\section{Mean field theory for the dipole condensate} \label{app:effective_action}
		
		In this appendix we use mean field theory to estimate the critical hopping strength at which the transition from the Mott insulator to the dipole condensate (DC) occurs. 
		
		We proceed as in App. B of \cite{lake2022dipolar}. We start by writing the hopping term in the microscopic Hamiltonian \eqref{dbhm_ham} as 
		\be H_{hop} = -\sum_{i,j} b^\da_{i}b_{i+1} [\mca]_{ij} b_{j+1}^\da b_j,\ee 
		where the matrix $\mca$ is defined as
		\be [\mca]_{ij} =t_3 (\d_{j,i+1} + \d_{i,j+1}) + t_4 (\d_{i,j+2} + \d_{j,i+2}).\ee 
		To determine where the transition into the DC occurs, we decouple the hopping term in terms of dipole fields $D_i$ as 
		\be H_{hop} = -\sum_{i} \( b_i^\da b_{i+1} D_i + (D_i)^\da b^\da_{i+1} b_i\) + \sum_{i,j} (D_i)^\da [\mca]\inv_{ij} D_j.\ee 
		We then integrate out the boson fields $b_i$ to produce an effective action for the $D$ variables. We will only be interested in obtaining the effective action to quadratic order in $D$ and derivatives thereof, which we parametrize as 
		\bea \label{app_dact} S_2 & = \int d\tau \, dx\,  \(w  |\p_\tau D|^2 + K_D |\p_x D|^2 + r |D|^2 \).\eea 
		In terms of $b_i$ correlation functions, perturbation theory yields 
		\bea S_2 & =  -  \frac1{2}\int d\tau_1\, d\tau_2\, \lan \mct [ H_{Db}(\tau_1) H_{Db}(\tau_2)]\ran + \int d\tau \sum_{i,j} (D_i)^\da [\mca]_{ij}\inv D_j ,\eea 
		with $H_{Db} \equiv - \sum_{i} b_i^\da b_{i+1} D_i + h.c$, and where the expectation value above is taken with respect to the ground state of the site-diagonal Mott insulating Hamiltonian 
		\be H_{onsite} =\frac12 \sum_i  \(Un_i(n_i-1) + U'n_in_{i+1}\).\ee 
		In what follows we will assume that at $t=0$, the system realizes a Mott insulator with $n>0$ bosons per site, whose ground state we write as $\k{\mi_n} \equiv \bigotimes_i |n\ran_i$.
		
		The first term in $S_2$ is calculated as 
		\bea \int\frac{d\o}\twp \sum_i &|D_i(\o)|^2\int d\tau\,  e^{i\o \tau}\, \lan T[(b^\da_i b_{i+1})(\tau) (b_i b^\da_{i+1})(0)]\ran \\ 
		& = \int\frac{d\o}\twp \sum_i |D_i(\o)|^2\int d\tau\, e^{i\o \tau}\sum_l \( \ct(\tau) e^{-\tau (E_l -E_0)} |\lan \mi_n| b^\da_i b_{i+1} |l\ran |^2 + \ct(-\tau)  e^{\tau (E_l - E_0)} |\lan \mi_n|b_{i+1}^\da b_i |l\ran |^2 \) \\ 
		& = \int \frac{d\o}\twp \sum_i |D_i(\o)|^2\sum_l |\lan \mi_n|b^\da_i b_{i+1} |l\ran|^2 \( \frac1{i\o + E_l - E_0} + \frac1{-i\o + E_l - E_0} \),
		\eea 
		where $E_0$ is the ground state energy of $H_{onsite}$ and $l$ runs over all of $H_{onsite}$'s eigenstates. 
		The only nonzero terms in the sum have 
		\be |\lan \mi_n|b_{i+1}^\da b_i |l\ran |^2  = n(n+1),\qq E_l - E_0 = U-\frac{U'}2,\ee 
		and so we may expand in small $\o$ and write 
		\bea \label{omegapart} \int_\o \sum_i |D_i(\o)|^2\int d\tau\, & e^{i\o \tau}\, \lan T[(b^\da_i b_{i+1})(\tau) (b_i b^\da_{i+1})(0)]\ran  = \frac{2n(n+1)}{U-U'/2} \int \frac{d\o}\twp \sum_i |D_i(\o)|^2 \( 1 - \frac{\o^2}{(U-U'/2)^2} \).\eea 
		This determines the coefficient $w$ of the time derivative term appearing in \eqref{app_dact} as 
		\be w = \frac{n(n+1)}{(U-U'/2)^3}.\ee 
		Note that as anticipated in \eqref{app_dact} no linear time derivative term of the form $D^\da \p_\tau D$ appears, due to the fact that spatial reflection acts as a particle-hole symmetry on the dipoles. 
		
		To derive $r$ and $K_D$, we write $\mca\inv \mca = \unit$ as 
		\be t_3 ([\mca\inv]_{i,j-1}+[\mca\inv]_{i,j+1}) + t_4([\mca]\inv_{i,j+2} + [\mca]\inv_{i,j-2}) = \d_{i,j},\ee 
		which tells us that 
		\be\sum_{jl} [\mca]\inv_{jl} e^{i(pj-ql)} = \frac{\d_{p,q}}{2(t_3 \cos(q) + t_4\cos(2q))}.\ee 
		Expanding in small $q$ and using the $\o$-independent part of \eqref{omegapart}, we obtain 
		\be r = \frac1{2(t_3+t_4)} - \frac{n(n+1)}{U-U'/2}.\ee 
		The mean-field transition thus occurs when 
		\be t_3 + t_4 = \frac{U-U'/2}{2n(n+1)},\ee
		so that at fixed $U,U'$ the transition occurs at a hopping strength that scales with $n$ as $1/n^2$. This estimate turns out to be in remarkably good agreement with numerics; see Fig. \ref{fig:dmrg}. 
		
		If we use the expressions for $t_3,t_4$ as derived in App. \ref{app:hopping}, the transition is estimated to occur at a single-particle hopping strength of 
		\be t_{sp} = V \sqrt{ \frac{1-U'/2U}{2(1+2n^2 +n)}}.\ee 
		Thus the presence of the nearest-neighbor repulsion and a large average density $n$ both help to push the transition down to smaller values of $t_{sp}$.

	\end{widetext} 
	
\end{document}